\documentclass[amssymb,twocolumn,aps]{revtex4}
\usepackage{graphics,color,graphicx,amsmath}
\usepackage{subcaption}
\usepackage{natbib}
\usepackage{verbatim}
\usepackage{graphicx}
\usepackage[above,below]{placeins}
\usepackage{float}
\usepackage{tabularx}
\usepackage{times}
\usepackage{blindtext}
\usepackage{url}
\usepackage{rotating}
\usepackage{mathtools}
\usepackage{threeparttable}
\usepackage{enumitem}

\begin{document}

\title{Beyond named methods: A typology of active learning based on classroom observation networks}

%Beyond named methods:

%The practices behind the name: A typology of active learning instruction based on classroom observation networks

\author{Meagan Sundstrom~\footnote[1]{sundstrommeagan@gmail.com}$^{1}$, Justin Gambrell$^{2}$, Colin Green$^{3}$, Adrienne L. Traxler$^{4}$, and Eric Brewe~\footnote[2]{Contact author: eb573@drexel.edu}$^1$}
\affiliation{$^1$Department of Physics, Drexel University, Philadelphia, Pennsylvania 19104, USA\\
$^{2}$Department of Computational Mathematics, Science and Engineering, Michigan State University, East Lansing, Michigan 48824, USA\\
$^3$Department of Physics, Bryn Mawr College, Bryn Mawr, Pennsylvania 19010, USA\\
$^4$Department of Science Education, University of Copenhagen, Copenhagen, Denmark}

\date{\today}

\begin{abstract}
A growing number of introductory physics instructors are implementing active learning methods in their classrooms, and they are modifying the methods to fit their local instructional contexts. However, we lack a detailed framework for describing the range of what these instructor adaptations of active learning methods look like in practice. Existing studies apply structured protocols to classroom observations and report descriptive statistics, but this approach overlooks the complex nature of instruction. In this study, we apply network analysis to classroom observations to define a typology of active learning that considers the temporal and interactional nature of instructional practices. We use video data from 30 instructors at 27 institutions who implemented one of the following named active learning methods in their introductory physics or astronomy course: Investigative Science Learning Environment (ISLE), Peer Instruction, Tutorials, and Student-Centered Active Learning Environment with Upside-down Pedagogies (SCALE-UP). We identify five types of active learning instruction: clicker lecture, dialogic clicker lecture, dialogic lecture with short groupwork activities, short groupwork activities, and long groupwork activities. We find no significant relationship between these instruction types and the named active learning methods; instead, implementations of each of the four methods are spread across different instruction types. This result prompts a shift in the way we think and talk about active learning: the names of developed active learning methods may not actually reflect the specific activities that happen during instruction. We also find that student conceptual learning does not vary across the identified instruction types, suggesting that instructors may be flexible when modifying these methods without sacrificing effectiveness. We discuss the implications of these results for both research and professional development of college physics instructors.
\end{abstract}

\maketitle

\section{Introduction}

An abundance of research demonstrates that active learning methods, where students engage in interactions with their peers and co-construct their understanding~\cite{piaget1926language,vygotsky1978mind}, promote conceptual development and improve the odds of student success more than traditional, lecture-based teaching methods in introductory level science courses~\cite{hake1998interactive,freeman2014active,theobald2020active}. Active learning encompasses a broad range of research-based instructional strategies, from specific activity types (e.g., computer simulations and small group work) to named methods (e.g., Peer Instruction~\cite{mazur1997peer,crouch2007peer}) and fully developed course plans (e.g., Investigative Science Learning Environment~\cite{etkina2007investigative,etkina2015millikan}).

Despite documented barriers to adoption, such as limited time to learn about active learning methods, rigid classroom spaces, and lack of institutional support~\cite{henderson2007barriers,henderson2012use,dancy2012experiences,Affriyenni2025navigating}, there is evidence that a growing number of physics instructors use active learning methods~\cite{foote2014diffusion,henderson2009impact,dancy2024physics}. As these methods become more widely implemented, instructors are continually modifying them to suit their specific contexts (e. g., student populations, institutional norms, and physical classroom spaces)~\cite{borrego2013fidelity,dancy2016implement,scanlon2019method}. These adaptations mean that active learning methods are rarely implemented `as is,' rather they take on different forms in practice.

Existing research, however, does not adequately capture the extent of or complexity within the variation in these different adaptations of active learning. Many studies rely on self-reported data from surveys and interviews of instructors about what they do during class~\cite{henderson2009impact,dancy2024physics,borrego2013fidelity,andrews2011active,scanlon2019method}, which may not be accurate~\cite{ebert2011we,dancy2016implement}. Other studies use direct classroom observation to examine instructional practices~\cite{turpen2009not,smith2014campus,wood2016characterizing,commeford2021characterizing,bukola2025perc}, but this work is largely descriptive (e.g., documenting the time spent on different activities) and overlooks nuances in instructional practices (e.g., activities happening simultaneously and the chronological order of activities, described further in the next section). We lack a detailed framework for describing the range of instructional practices that occur in active learning classrooms. The main goal of this article, therefore, is to characterize different implementations of well-established active learning methods in introductory physics. We analyze 114 h of classroom video recordings from 30 instructors at 27 different institutions using classroom observation networks, which capture the temporal and interactional nature of classroom activities and move beyond descriptive statistics, to define a typology of active learning instruction.

The specific ways that active learning methods are implemented may also relate to their effectiveness in supporting student outcomes, including conceptual learning~\cite{weir2019small,bukola2025perc,sundstrom2025relativebenefits,cleveland2017investigating,connell2016increasing,mcneal2020biosensors}. Weir and colleagues~\cite{weir2019small}, for example, find that in introductory biology, certain types of activities (e.g., worksheets) are more strongly associated with gains in student conceptual understanding than others (e.g., clicker questions). Our previous work suggests that different implementations of Peer Instruction in introductory physics (specifically, the extent to which activities are fully interactive, such as clicker questions, versus vicariously interactive, such as listening to a peer respond to an instructor question) may be differentially associated with student learning~\cite{bukola2025perc}. Another aim of this article, therefore, is to relate different types of active learning instruction to student conceptual learning.

%In the remainder of this paper, we describe existing literature related to classroom observations, outline our methods for data collection and analysis, present the results, and discuss the implications of this work.

\section{Background and Current Study}

Here we summarize how researchers have conducted and analyzed classroom observations in prior work and outline the approaches used in the current study.

\subsection{Conducting classroom observations}

Systematic classroom observations are typically conducted using established protocols, or tools used to reliably document student and instructor activities during instruction. In their literature review~\cite{anwar2021systematic}, Anwar and Menekse identify eight classroom observation protocols that have been used to characterize instructional practices in college-level science courses: Reformed Teaching Observation Protocol (RTOP)~\cite{sawada2002measuring}, Oregon Teacher Observation Protocol (OTOP)~\cite{wainwright2003development}, VaNTH Observation System (VOS)~\cite{harris2003developing}, Cooperative Learning Observation Protocol (CLOP)~\cite{kern2007cooperative}, Teaching Dimension Observation Protocol (TDOP)~\cite{hora2013instructional,hora2015toward}, Classroom Observation Protocol for the Undergraduate STEM (COPUS)~\cite{smith2013classroom}, Classroom Interactive Engagement Observation Protocol (CIEOP)~\cite{kothiyal2013effect}, and Student Resistance and Instructional Practices (StRIP)~\cite{finelli2014classroom}. These protocols span both open-ended (e.g., writing field notes) and structured (e.g., checking off when pre-defined activities occur) approaches, use Likert-scale ratings of activities and/or record whether or not pre-defined activities occur (without rating the activities), focus on student and/or instructor behaviors, and can be conducted in real time (i.e., with a live observer in the classroom) and/or with classroom video recordings.

In this study, we use the COPUS, which records the presence of pre-defined classroom activities at two-minute time intervals (described further in Sec.~\ref{copusmethods}). We chose this protocol because it was designed to identify the range and frequency of instructional practices, in line with the goals of our analysis~\cite{smith2013classroom,anwar2021systematic}. Furthermore, the COPUS is both structured (i.e., not open-ended or ratings-based, which may be biased by an observer's subjective viewpoints) and standardized (i.e., it can be compared across different courses). The COPUS also distinguishes student and instructor activities, allowing our analysis to only focus on the instructor portion of the protocol (i.e., to characterize instructional practices). Finally, the COPUS is suitable for either real-time observations or video recorded observations. We collected and analyzed video recordings of classroom instruction so that we could include data from many institutions and iteratively establish reliability among multiple observers.

\subsection{Analyzing classroom observations}

Once a structured, standardized classroom observation protocol is applied in real time or to a video recording, researchers typically conduct quantitative analyses to understand the types of activities that occurred during instruction. The simplest form of analysis relies on descriptive statistics, such as calculating the fraction of class time spent on each activity in the protocol. A handful of studies have used descriptive statistics to characterize instruction~\cite{turpen2009not,smith2014campus,wood2016characterizing,commeford2021characterizing,bukola2025perc}; for example, Commeford and colleagues~\cite{commeford2021characterizing} compare the fractions of class time spent on each activity in the COPUS across six named active learning methods in introductory physics. Such descriptive studies are useful for documenting prevalent classroom activities and establishing baselines of instructional practices, but they are not scalable: they are not suited to compare a large number of classroom observations.

To address this scalability issue, another body of work has applied clustering algorithms, such as \textit{k}-means clustering, latent class analysis, and latent profile analysis (LPA), to the descriptive statistics of many observations in order to identify groups of observations with similar features~\cite{lund2015best,denaro2021comparison,wan2020characterizing,campbell2017comprehensive,stains2018anatomy,commeford2022characterizing,weston2023measures,sundstrom2025relativebenefits}. Across many COPUS observations of introductory level science courses, for example, Stains and colleagues use LPA to identify three types of instruction: didactic, interactive, and student-centered~\cite{stains2018anatomy}. Our recent work applies LPA to COPUS observations of four named active learning methods in introductory physics~\cite{sundstrom2025relativebenefits}. We find four types or ``profiles" of instruction, each of which is characterized by one predominant activity: lecture, clickers (i.e., clicker questions), worksheets, and other groupwork (which encompasses, for example, laboratory activities and solving problems at whiteboards). 

These clustering methods, however, still make use of descriptive statistics related to classroom instruction, and, as Erdmann and Stains note, ``the use of summary statistics alone has the potential to obscure interesting trends and meaningful differences" in instructional practices~\cite{stains2019genome} (p. 2). Descriptive statistics ignore the interactional and chronological nature of classroom activities: multiple activities may occur simultaneously and activities precede and follow one another over time rather than occur in isolation. Aggregating the total time spent on an activity also masks any information about the duration of each instance of the activity. As an example, consider two instructors of an hour-long class whose descriptive statistics indicate that they both guided students through a worksheet activity for 40 min and led a follow-up discussion of that activity for 20 min. These two instructors' classes may have been looked very different in practice: one may have proctored the worksheet activity for 40 minutes and then followed up on that worksheet for 20 min, while the other may have cycled through proctoring worksheets for 10 min and then following up for 5 min four times. There remains a need for an analysis tool that captures these subtleties in instructional practices.

Erdmann and Stains outline one possible method to advance our analysis of classroom observations~\cite{stains2019genome}. The authors apply methods from bioinformatics, treating classrooms as genomes. Essentially, they represent COPUS observations as strings of activities, where activities and their durations are represented along a timeline (indicating their order) and activities can overlap with one another. This method overcomes some of the weaknesses of descriptive statistics mentioned above: it captures the sequential nature of instructional events as well as co-occurrence. The approach can also be scaled to look at patterns across many observations, such as identifying the fraction of all observations in which different activities precede and follow clicker questions to get a better sense of how clicker questions are implemented. One limitation of this method, however, is that it does not facilitate quantitative comparisons between observations at the grain size of an entire observation (i.e., the example in the previous sentence only focuses on one activity, clicker questions). To do so, a researcher would need to qualitatively examine the complete strings of activities across observations and note similarities and differences.

Hora and colleagues introduce yet another approach to capture complexities in instructional practices: classroom observation networks~\cite{hora2013instructional,hora2014remeasuring,hora2015toward}. In their analysis of TDOP observations, network nodes represent activities from the protocol and network edges represent the relative amount of time that pairs of activities co-occur. The networks contain information about both time duration of activities and co-occurrence, while also affording direct comparisons between observations (either between different observations from a single instructor or between observations from different instructors) because every network has the same set of nodes and edges can be normalized by class duration. For instance, in one of their studies~\cite{hora2014remeasuring}, the researchers compare the classroom observation networks of two different biology instructors whose instructional styles appeared similar according to the descriptive statistics alone: they both lectured for most of the class time. The networks, however, distinguished specific features of each instructor's lecturing style. One instructor typically used Powerpoint slides and integrated anecdotes and humor into their lectures, while the other instructor integrated various student-centered activities and instructional technologies into their lectures.

Classroom observation networks, therefore, offer a useful tool for both capturing nuance in instructional practices \textit{and} directly comparing different observations to one another. We leverage this network approach in the current study to categorize the teaching practices of the 30 instructors in our dataset. 

\subsection{Current study}

Classroom observation networks have the potential to improve our characterization of active learning methods, but to our knowledge they have not been applied in other contexts beyond the few studies in biology mentioned above~\cite{hora2013instructional,hora2014remeasuring,hora2015toward}. In the current study, we use such networks to explore variation in introductory physics instructors' implementations of named active learning methods. To define a typology of active learning instruction (i.e., identify and interpret groups of instructors with similar teaching practices), we move beyond Hora and colleagues' qualitative comparison of network structures and quantify the extent to which instructors' observation networks are similar using cosine similarity (described further in the next section). We then follow the ``network-of-networks" approach established by Bruun and colleagues, where the grouping of similar classroom observation networks is performed by applying a clustering algorithm to a similarity network in which the nodes represent the classroom observation networks themselves and the edges are the cosine similarities between the networks~\cite{bruun2025network}. This approach allows us to not only distinguish and characterize the instruction types, but also to visualize the relationships \textit{between} the types (which is not possible with other clustering methods, such as \textit{k}-means clustering, latent class analysis, and latent profile analysis). We also examine the relationships between instruction type, course and institution attributes, and student conceptual learning. The following research questions guided our study:
\begin{enumerate}
    \item What typology of instructional practices used in named active learning methods may be inferred from the resultant clusters of classroom observation networks?
    \item How are the instruction types related to the profiles found in our previous analysis of these data~\cite{sundstrom2025relativebenefits}, named active learning method, and other course and institution attributes (e.g., number of enrolled students)?
    \item How are the instruction types related to student conceptual learning?
\end{enumerate}

\section{Methods}

In this section, we first describe our procedures for collecting classroom observation and student conceptual learning data and then provide step-by-step details on how we analyzed these data. All analysis was conducted in R (Version 4.3.2). De-identified data and analysis scripts are available at Ref.~\cite{github2025}.

\subsection{Data collection}

As part of an ongoing national research project aimed at characterizing different active learning methods used to teach introductory physics, we recruited instructors who self-reported using one of the following active learning methods in their introductory physics or astronomy course: 
\begin{enumerate}
    \itemsep0cm
    \item Investigative Science Learning Environment (ISLE)~\cite{etkina2007investigative,etkina2015millikan}: In all course components (``whole-class" implementation), or only in laboratories (``lab-only" implementation), students take on the role of novice scientists and engage in a lab-based learning cycle. In small groups, students observe a physics experiment, explain their observations, make predictions about new experiments, design and conduct these experiments, and revise their explanations. 
    \item Peer Instruction~\cite{mazur1997peer,crouch2007peer}: During lectures, students work in small groups of nearby peers to answer multiple choice questions using electronic clickers or another form of voting system. Typically, class sessions start with a short lecture and then cycle through several multiple choice questions where
    the instructor poses a question, students answer the question individually, students discuss the question in small groups, students re-answer the question, and then the instructor explains the solution. 
    \item Tutorials for introductory physics~\cite{mcdermott2002tutorials} and astronomy~\cite{adams2003lecture}: In all course components (``whole-class" implementation), or only in recitation sections (``recitation-only" implementation), students complete worksheets in small groups. These worksheets are highly scaffolded and are intended to elicit, confront, and resolve common misconceptions.
    \item Student-Centered Active Learning Environment with Upside-down Pedagogies (SCALE-UP)~\cite{beichner2007student}: 
    The course components (i.e., lectures, labs, and recitations) are fully integrated and take place in a classroom designed to promote student interaction, often containing large tables that seat nine students and whiteboards along the classroom perimeter. During class sessions, students solve conceptual and quantitative physics problems and complete lab activities in small groups.
\end{enumerate}
We note that several other named active learning methods for introductory physics exist and were part of the preliminary study for this project~\cite{commeford2021characterizing, commeford2022characterizing} (e.g., Modeling Instruction~\cite{brewe2008modeling} and Context-Rich Problems~\cite{heller1992teaching,heller1992teaching2}). We limited our current study to the above four methods because they are both well-known and widely implemented~\cite{dancy2024physics}, and we aimed to collect data from multiple implementations of each method.

We recruited instructors in several ways. First, we contacted a grant advisory board member affiliated with each active learning method and asked for names of instructors who were known to be using the method. Beyond these lists, we used additional approaches such as reaching out to instructors in our personal networks, posting advertisements for our study within the American Physical Society, and snowball sampling (i.e., asking instructors who participated in our study to identify other instructors they knew who were implementing the same method). Data collection took place in the Fall 2023, Spring 2024, and Fall 2024 semesters. In total, 31 instructors participated in the study, each with one course in one semester (i.e., no instructors participated in the study with multiple courses). Each instructor received \$1,000 for participating. 

In this paper, we focus on two data sources that we asked participating instructors to collect: classroom observations and concept inventories (i.e., research-developed and validated assessments used to measure student understanding of physics concepts). 30 out of the 31 instructors collected classroom observations and 26 out of these 30 instructors collected concept inventory data. All 30 of these instructors are included in the analysis presented here (Table~\ref{tab:metadata}). These instructors come from 27 unique institutions in the United States, including both public and private institutions, a range of research-intensive and undergraduate-focused institutions, and several minority-serving institutions (Table~\ref{tab:institutioninfo}).

\begin{table}[t] 
\centering
\caption{\label{tab:metadata}
Number of courses included in the study by data source and by active learning method. Courses included in the concept inventory analysis are a subset of the courses in the classroom observation analysis.}
\begin{ruledtabular}
\setlength{\extrarowheight}{1pt}
\begin{tabular}{lccccc}
Method &  Observations & Concept Inventory  \\
\hline
ISLE   & 6  & 5\\
Peer Instruction  & 9  & 9\\
Tutorials&  8& 7\\
SCALE-UP   & 7 & 5\\
\end{tabular} 
\end{ruledtabular}
\end{table} 

\begin{table}[t] 
\centering
\caption{\label{tab:institutioninfo}
Characteristics of the 27 unique institutions included in our study. Carnegie Classifications of Research Activity are from 2025. R1 indicates ``Very High Research Spending and Doctorate Production," R2 indicates ``High Research Spending and Doctorate Production," and RCU indicates ``Research Colleges and Universities."}
\begin{ruledtabular}
\setlength{\extrarowheight}{1pt}
\begin{tabular}{lc}
Type of Institution &  \textit{N} (\%)\\ 
\hline
Public/private &  16 (59\%)/11 (41\%)\\ 
Highest degree awarded\\
\hspace{2mm} Associate's & 1 (4\%)\\ 
\hspace{2mm} Bachelor's & 2 (7\%)\\ 
\hspace{2mm} Master's & 9 (33\%)\\ 
\hspace{2mm} PhD & 15 (56\%)\\ 
Carnegie Classification of Research Activity &  \\ 
\hspace{2mm} R1 & 10 (37\%)\\ 
\hspace{2mm} R2 & 3 (11\%)\\ 
\hspace{2mm} RCU & 5 (19\%)\\ 
Minority-serving status &  \\ 
\hspace{2mm} Hispanic-Serving Institution & 3 (11\%)\\ 
\hspace{2mm} Women's College & 1 (4\%)\\ 
\end{tabular} 
\end{ruledtabular}
\end{table}

\begin{table*}[t] 
\centering
\caption{\label{tab:codes}
Abbreviations and definitions of the nine instructor codes in the COPUS~\cite{smith2013classroom} that we included in this study.}
\begin{ruledtabular}
\setlength{\extrarowheight}{1pt}
\begin{tabular}{lll}
Code & Abbreviation & Definition  \\
\hline
Lecturing & Lec  & Presenting content to the whole class  \\
Real-time Writing & RtW  & Writing in real time on a projector or board while presenting to the whole class \\
Clicker Question & CQ  & Asking a clicker (or other voting system) question to the students \\
Posing Question & PQ  & Asking a non-clicker (and non-rhetorical) question to the students \\
Moving and Guiding & MG  & Moving through class guiding ongoing student work during an active learning task  \\
Following Up & FUp  & Providing feedback on clicker question or other activity to the whole class  \\
Answering Question & AnQ  & Listening to and answering student questions in front of the whole class  \\
Demonstrations/Videos & D/V  & Showing a demonstration, video, simulation, or animation  \\
Administration & Adm  & Presenting administrative information (e.g., related to exams) to the whole class\\
\end{tabular} 
\end{ruledtabular}
\end{table*} 

For the classroom observations, we asked instructors to use Zoom, their own camera, or a camera that we provided to them via mail to record videos of three consecutive class sessions, capturing a ``typical week" in the course as recommended by prior studies~\cite{weir2019small, stains2018anatomy}. All courses were conducted in person and videos included both students and the instructor(s) in the camera's view. All 30 instructors recorded three full class sessions, totaling 114 h of classroom video. In all ISLE courses, the instructors only recorded the lecture sections (i.e., not lab sections for the lab-only implementations, where the method was actually implemented). In all Peer Instruction courses, instructors only recorded lecture sections (and not any lab or recitation sections student may have also been attending). In one recitation-only Tutorials course, the instructor only recorded lecture sections (and not recitation sections, where the method was actually implemented). In the other Tutorials courses, instructors recorded the class section where Tutorials were being implemented (i.e., lecture sections for whole-class implementations and recitation sections for recitation-only implementations). All SCALE-UP instructors recorded the entirety of three class sessions (i.e., possibly including lab activities). We denote these different cases when we present the results.

For the concept inventories, we asked instructors to choose an existing concept inventory to give to their students, such that the concept inventory aligned with the topics taught in the course (e.g., because instructors taught a range of student populations and courses spanned both physics and astronomy). Instructors administered the assessment at both the beginning (before any exposure to the material) and the end of the semester. Most instructors used the Force Concept Inventory~\cite{hestenes1992force}, Force and Motion Concept Evaluation~\cite{ramlo2008validity}, or Mechanics Baseline Test~\cite{hestenes1992mechanics} (Table~\ref{tab:individualfeatures} in the Appendix). All 26 instructors had more than 40\% of and/or more than 30 enrolled students with matched responses (i.e., completing both the pre- and post-semester concept inventory; response rates ranged from 41\% to 98\%, with a mean of 72\%). We did not impute any concept inventory data because we did not have any information about individual students (e.g., demographics) besides their scores~\cite{nissen2019missing}.

\subsection{Data analysis}

\subsubsection{Video coding}
\label{copusmethods}

We used the COPUS~\cite{smith2013classroom} to characterize instruction from the classroom video recordings. The COPUS records whether or not certain student and instructor activities, or ``codes," occur in each two-minute time interval.

The first three authors individually applied the COPUS to one video observation (i.e., one of the three class sessions recorded by one instructor) from each of the four active learning methods. We then met to discuss disagreements and clarify inclusion and exclusion criteria for each COPUS code. The three coders iteratively re-coded the videos and met for discussions until we reached greater than 80\% agreement (as measured with Cohen's Kappa~\cite{landis1977measurement}) for each of the four observations for each pair of coders. After reaching this level of agreement, we randomly assigned one coder to each of the remaining observations such that any resulting patterns among the observations are not attributable to systematic differences between coders. The first author coded 46 video observations, the second author coded 32 video observations, and the third author coded 12 video observations.

In our coding, we used the full COPUS, which includes 13 student codes and 12 instructor codes. We only included instructor codes in the analysis presented here, however, as we aimed to characterize instructional practices (Table~\ref{tab:codes}). We excluded three of the instructor codes (one-on-one extended discussion with one or a few students while ignoring the rest of the class, waiting around, and other) because they were rarely applied to the observations in our dataset (each of these three codes was observed in less than 2\% of all two-minute time intervals across the entire dataset).

\subsubsection{Constructing classroom observation networks}

We constructed one classroom observation network for each of the 30 instructors (i.e., we aggregated the three video observations together for each instructor). All networks included nine nodes, one per COPUS code (Table~\ref{tab:codes}). Directed edges represented the chronological order of codes, pointing from an initial code to another code that occurred subsequently in the observation. The edges were also weighted as the number of transitions that occurred from one code to another divided by the total number of two-minute time intervals in the instructor's three recordings (to normalize for different class lengths across instructors). We focused on transitions between codes rather than the co-occurrence of codes, as in Hora and colleagues' work~\cite{hora2013instructional,hora2014remeasuring,hora2015toward}, because instructor codes do not frequently co-occur (i.e., the instructor predominantly does one activity at a time) and we wanted to capture the flow of classroom activities over time (as in Ref.~\cite{stains2019genome}).

\begin{figure}[t]
    \centering
    \includegraphics[width=3in,trim={3cm 3.5cm 2cm 2.7cm}]{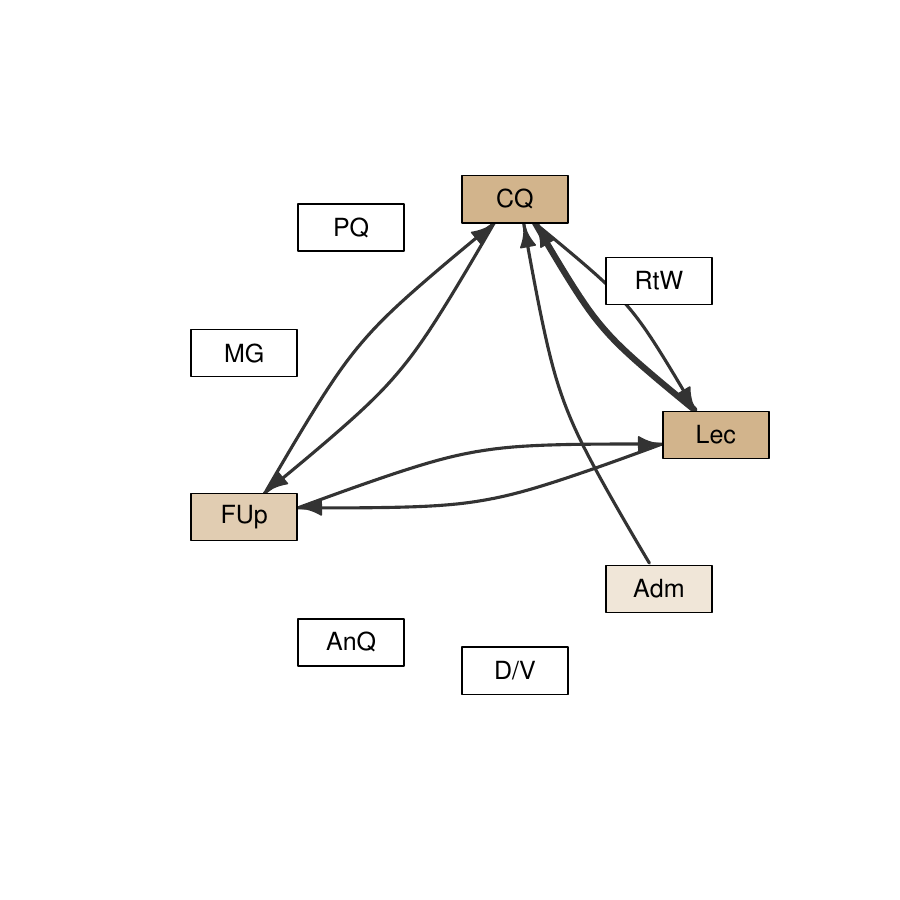}
    \caption{Toy classroom observation network for the example COPUS coding in Table~\ref{tab:exampleCOPUS}.  Though not directly part of our analysis, node color indicates the fraction of all two-minute time intervals that the code was present: darker colors indicate larger fractions (i.e., longer time durations). Edges point from an initial code to the code that occurs in the following two-minute time interval of the COPUS observation. Edge width indicates the number of transitions occurring between those two codes normalized by the total number of two-minute time intervals observed.
    }
    \label{fig:toynetwork}
\end{figure}

The chronological order of codes is not obvious in COPUS observations because multiple codes can be recorded in the same time interval. If an observer records that two codes occur in the same two-minute time interval, this could indicate either that the two codes occurred simultaneously or that one code occurred and then the other code occurred afterward (and we cannot tell which one occurred first). As our analysis aimed to characterize transitions between instructional activities and not the co-occurrence of activities, we only considered a transition between codes to be when a code that was not present in the previous two-minute time interval became present in the following two-minute time interval. We created an edge from all codes in the previous time interval to the newly appearing code(s) in the following interval. As an example, consider the ten minutes of COPUS coding in Table~\ref{tab:exampleCOPUS}. This observation would yield the following edges in an observation network (shown in Fig.~\ref{fig:toynetwork}):
\begin{itemize}[nosep]
    \item Lec $\rightarrow$ CQ (weight = 2/5 = 0.4), from the first to second interval and from the fourth to fifth interval,
    \item Adm $\rightarrow$ CQ (weight = 1/5 = 0.2), from the first to second interval,
     \item Lec $\rightarrow$ FUp (weight = 1/5 = 0.2), from the second to third interval,
    \item CQ $\rightarrow$ FUp (weight = 1/5 = 0.2), from the second to third interval,
    \item CQ $\rightarrow$ Lec (weight = 1/5 = 0.2), from the third to fourth interval,
    \item FUp $\rightarrow$ Lec (weight = 1/5 = 0.2), from the third to fourth interval, and
    \item FUp $\rightarrow$ CQ (weight = 1/5 = 0.2), from the fourth to fifth interval. 
\end{itemize}
The first two intervals in Table~\ref{tab:exampleCOPUS}, for example, do not produce an edge from Lec to Lec or from Adm to Lec because Lec occurs in the first interval and so may have already been happening when the second time interval started, which would not be a transition to a new code. Our definition of transitions between codes also prevents any self-loops (i.e., edges from a node to itself, indicating the instructor is doing the same activity in multiple back-to-back time intervals) from appearing in the network. In cases where the instructor is doing the same activity for long periods of time, these self-loops would dominate the network and information about transitions between activities would be suppressed. We note, though, that our observation networks still reflect whether individual codes occurred for long, continuous time periods versus short, interspersed time periods. If a network only has a few edges (i.e., few transitions between codes), for example, this would imply that the instructor was doing only a few activities for extended time periods.

\begin{table}[t] 
\centering
\caption{\label{tab:exampleCOPUS}
Toy example of COPUS coding for ten minutes of class time. Each row represents a two-minute time interval. One (zero) indicates the presence (absence) of the code within the two-minute time interval.}
\begin{ruledtabular}
\setlength{\extrarowheight}{1pt}
\begin{tabular}{lcccccccc}
Lec & RtW & CQ & PQ & MG & FUp & AnQ & D/V & Adm \\
\hline
1 &  0 & 0 &0&0&0&0&0& 1 \\
1 & 0  &  1&0&0&0&0&0&0  \\
 0& 0  &  1&0&0&1&0&0&0  \\
  1&   0&0  &0&0&1&0&0&0  \\
   0&  0 &  1&0&0&0&0&0&0  \\
\end{tabular} 
\end{ruledtabular}
\end{table}

\subsubsection{Constructing the similarity network}

To quantify the extent to which each pair of classroom observation networks was similar, we considered two measures: the Jaccard index (as in Ref.~\cite{dalka2024network}) and cosine similarity (as in Refs.~\cite{tapping2018visualizing,kortemeyer2014extending,odden2020thematic}). Both measures have been used to compare networks with the same set of nodes (as we have in our study)~\cite{tantardini2019comparing} and have been shown to produce similar rankings of similarity between pairs of networks~\cite{steinert2016comparative}.

The Jaccard index measures specific edge agreement, or the amount of overlap between the edges present in two networks. For a pair of directed, weighted networks $G_1$ and $G_2$ with the same set of nodes $V$ and respective sets of edges $E_1$ and $E_2$ and edge weights $W_1$ and $W_2$, the Jaccard index is calculated as: 
\begin{equation*}
    J(G_1,G_2) = \frac{\sum\limits_{e \in E_1 \cap E_2} \min(W_1(e), W_2(e))}{\sum\limits_{e \in E_1 \cup E_2} \max(W_1(e), W_2(e))}.
\end{equation*}
Values range from 0 to 1, with values closer to 1 indicating that the two networks have more similar sets of edges and corresponding edge weights.

Cosine similarity, on the other hand, measures the structural similarity between two networks. For a pair of directed, weighted networks $G_1$ and $G_2$ with the same set of nodes $V$ and respective vectors of edge weights for each pair of nodes $\vec{v}_1 = [W_1(1,2), W_1(1,3),...,W_1(V,V-1)]$ and $\vec{v}_2 = [W_2(1,2), W_2(1,3),...,W_2(V,V-1)]$, cosine similarity is calculated as: 
\begin{equation*}
    \text{cosine similarity} = \frac{\vec{v}_1 \cdot \vec{v}_2}{||\vec{v}_1 || \ ||\vec{v}_2 ||}.
\end{equation*}
Values range from 0 to 1 (assuming there are no negative edge weights, as is the case in our analysis), with values closer to 1 indicating that the two networks have more similar sets of edges and patterns in the relative distribution of edge weights.

We calculated both of these  measures for every pair of classroom observation networks in our analysis (Fig.~\ref{fig:Heatmap} in the Appendix). Cosine similarity produced many more intermediate values in its possible range, whereas the Jaccard index was suppressed to low values for almost every pair of classroom observation networks (previous work has also shown that this measure is volatile when networks are sparse~\cite{walsh2020connecting,sundstrom2022examining}). We used cosine similarity in the remainder of our analysis because the wider range of values allows for more distinct subgroups of observation networks to be created.

The matrix of cosine similarity values between every pair of classroom observation networks can itself be represented as a network, and we will refer this network as the ``similarity network." In the full similarity network, there are 30 nodes representing the classroom observation networks (one per instructor) and 435 undirected edges (there is an edge connecting every possible pair of instructors) with weights equal to the cosine similarity between the two connected instructors.

\begin{table*}[t] 
\centering
\caption{\label{tab:networkdefs}
Node-level and network-level measures calculated for the classroom observation networks assigned to each cluster.}
\begin{ruledtabular}
\setlength{\extrarowheight}{1pt}
\begin{tabular}{l p{15.5cm}}
Measure & Definition  \\
\hline
Node-level & \\
\hspace{0.3cm}Indegree & The number of adjacent, incoming edges to a node (ignoring edge weight). In this study, the number of unique other COPUS codes that precede a given code.\\
\hspace{0.3cm}Outdegree & The number of adjacent, outgoing edges to a node (ignoring edge weight). In this study, the number of unique other COPUS codes that follow a given code.\\
%\hspace{0.5cm}Instrength & The sum of weights of adjacent, incoming edges to a node. In this study, the number of times any other COPUS code precedes a given code (normalized by class length).\\
%\hspace{0.5cm}Outstrength & The sum of weights of adjacent, outgoing edges to a node. In this study, the number of times any other COPUS code follows a given code  (normalized by class length).\\
\hspace{0.3cm}Betweenness & The fraction of all shortest paths between every pair of nodes in the network (where path length is the sum of edge weights) that pass through a given node. In this study, the extent to which a code serves as a ``bridge" between other codes. \\
Network-level & \\
\hspace{0.3cm}Edges & The total number of edges in the network (considering direction but ignoring edge weight). In this study, the number of unique pairs of COPUS codes with transitions occurring between them.\\
\hspace{0.3cm}Total strength & The sum of all edge weights in the network. In this study, the total number of transitions between COPUS codes (normalized by class length).\\
\hspace{0.3cm}Mutuality & The fraction of edges that are reciprocal (i.e., there are edges from code A to code B \textit{and} from code B to code A). In this study, the extent to which transitions between codes occur in both orders (e.g., code A both precedes and follows code B) rather than one (e.g., code A always precedes code B).\\
\hspace{0.3cm}Transitivity & The proportion of two-paths (e.g., code A follows code B and code B follows code C) that close to form triangles (e.g., code C also follows code A). In this study, the extent to which codes occur in short, cyclical sequences.\\
%\hspace{0.5cm}Diameter & Of the shortest paths between every pair of nodes in the network, the length of the longest of those paths. In this study, the extent to whih codes occur in a long, chain-like sequence.\\
\end{tabular} 
\end{ruledtabular}
\end{table*}

\subsubsection{Sparsifying the similarity network}

The fully connected similarity network is difficult to analyze; therefore, we followed the approach in Ref.~\cite{bruun2025network} which uses the locally adaptive network sparsification (LANS) algorithm~\cite{foti2011nonparametric} to remove insignificant edges. Rather than setting a universal, fixed threshold where we remove edges with weights lower than the threshold (which may remove edges with low weights compared to the rest of the network but that are locally important to the nodes to which they are connected), the LANS algorithm considers the distribution of edge weights for each individual node and preserves edges that are statistically significant for that node. Mathematically, the LANS algorithm makes use of fractional edge weights between nodes $i$ and $j$, defined as:
\begin{equation*}
    p_{ij} = \frac{W_{ij}}{\sum\limits_{k = 1}^{N_i} W_{ik}},
\end{equation*}
where $N_i$ is the number of nodes connected to node $i$. For each node in the network, the fraction of edges with weights less than or equal to $p_{ij}$ is:
\begin{equation*}
    \hat{F}(p_{ij}) = \frac{1}{N_i}\sum\limits_{k = 1}^{N_i} \textbf{1} \{ p_{ik} \le p_{ij} \},
\end{equation*}
where \textbf{1} is the indicator function (equal to 1 if the statement in brackets is true and 0 otherwise). For a given node, $\hat{F}(p_{ij})$ is effectively the probability of choosing an adjacent edge at random that has a weight less than or equal to $p_{ij}$. Therefore, edges are considered to be locally significant if $1 - \hat{F}(p_{ij})$ is less than a pre-determined significance level, $\alpha$. All non-significant edges are removed from the network, retaining only those edges that are locally important to the network structure. In this analysis, we set $\alpha$ = 0.05, which retained 46 edges in our similarity network and did not leave any nodes isolated (i.e., with no adjacent edges). A detailed description of the methods we used to choose this significance level is provided in the Appendix (Fig.~\ref{fig:Alpha}).

\subsubsection{Identifying clusters in the sparsified similarity network}

Following Ref.~\cite{bruun2025network}, we applied the Infomap community detection algorithm~\cite{rosvall2008maps,bruun2014time} to the sparsified similarity network to find clusters of nodes (i.e., classroom observation networks) that are more similar to each other than than they are to other nodes. The Infomap algorithm identifies clusters within a network by modeling the flow of information as a random walk and compressing the description of this flow, grouping nodes into clusters that minimize the expected description length of a random walker's path. The expected description length uses a two-level codebook: one for clusters and one for nodes within the clusters. Mathematically, the algorithm aims to minimize the map equation:
\begin{equation*}
    L(M) = q_{\curvearrowright} H(\mathcal{Q}) + \sum_{i=1}^{m} p_{\circlearrowright}^{i} H(\mathcal{P}^{i}),
\end{equation*}
where \( q_{\curvearrowright} \) is the probability of the random walker exiting a cluster, \( H(\mathcal{Q}) \) is the entropy of the cluster-level codebook, \( p_{\circlearrowright}^{i} \) is the probability of the random walker staying within cluster \( i \), \( H(\mathcal{P}^{i}) \) is the entropy of the codebook within cluster \( i \), and \( m \) is the number of clusters. Further mathematical details can be found in Ref.~\cite{rosvall2008maps}. 

The output of the algorithm is a partition of the similarity network into clusters of nodes (i.e., classroom observation networks) with similar characteristics, with each node assigned to one cluster. 

\subsubsection{Interpreting the clusters}

To define a typology of active learning instruction (i.e., the first research question), we characterized each emergent cluster using several node-level and network-level measures (summarized in Table~\ref{tab:networkdefs}) similar to prior work~\cite{lopez2025mining}. We calculated the average value of each of these measures for the classroom observation networks assigned to each cluster. Node-level measures were calculated for all nine nodes (i.e., COPUS codes) in the classroom observation networks and provided information about the role and relative importance of each COPUS code to each instructor's style (e.g., whether clickers are central to the way they teach). Network-level measures provided information about overall instructional style (e.g., distinguishing instructors who frequently transition between activities from those who do a few activities for extended periods of time). We triangulated these average measures with the classroom observation networks themselves and our qualitative knowledge of the classroom video recordings to inform our labels and interpretations of the clusters. These labels and interpretations were discussed and negotiated among the full research team until we reached consensus.

\subsubsection{Segregation analysis}

With the typology of active learning instruction defined, we aimed to identify the extent to which instructors were assigned to clusters (i.e., instruction types) with other instructors having similar course or institution attributes (i.e., the second research question). We examined the following attributes: 
\begin{itemize}[nosep]
    \item profile from our previous LPA~\cite{sundstrom2025relativebenefits} of these data to which the majority of the instructor's three classroom observations were assigned (lecture, clickers, worksheets, or other groupwork; the previous analysis did not aggregate each instructor's three class sessions; we exclude one course whose three observations were each assigned to different profiles),
    \item active learning method (ISLE, Peer Instruction, Tutorials, or SCALE-UP; we did not separate different implementations of ISLE and Tutorials given the small sample sizes),
    \item class size (smaller or larger than 45 enrolled students, see Fig.~\ref{fig:ClassSize} in the Appendix for details on how we chose this threshold),
    \item discipline (physics or astronomy), 
    \item public or private institution,
    \item research designation (R1/R2 or other), and
    \item PhD-granting or non-PhD-granting institution (see Table~\ref{tab:individualfeatures} in the Appendix).
\end{itemize}
We looked at the alignment of instruction types with our previous latent profile analysis to compare the grouping of classroom observations based on a more descriptive approach to that based on classroom observation networks. We investigated active learning method to understand the extent to which instructional practices vary within and across the named methods. We examined class size because previous research indicates that there are challenges to incorporating active learning strategies in larger classes; therefore, instructional practices may vary between small and large classes~\cite{murdoch2002active}. We included the last three variables because it is plausible that instructors at institutions that devote more resources to teaching than research may be supported in using different types of instruction than those at research-focused institutions.

For each of these attributes, we calculated the Segregation Z-score, established in Ref.~\cite{bruun2014time}, of the clustered similarity network. The authors of Ref.~\cite{bruun2014time} define a network-level Segregation measure as follows:
\begin{equation*}
    D_{seg}(M) = \frac{1}{N}\sum_{k \in M} N_k (\sum_{t=1}^{s} \ p_{kt} \text{log}_2(\frac{p_{kt}}{q_t})),
\end{equation*}
where $M$ is the clustering of the network (in our case, from the Infomap algorithm), $N$ is the number of nodes in the network, $s$ is the number of different categories or levels of the attribute of interest, $p_{kt} =N_{kt}/N_k $ is the probability of choosing a node at random from cluster $k$ that has attribute $t$, and $q_t = N_t / N$ is the probability of choosing a node at random from the entire network with attribute $t$. The Segregation Z-score compares the above Segregation measure for our similarity network to that for a large sample of networks $r$ (we used 10,000 networks as in Ref.~\cite{bruun2014time}) where the attributes are randomly distributed across the nodes while keeping the network structure and clustering the same:
\begin{equation*}
    Z = \frac{D_{seg}(M) - \langle D_{seg}^r(M) \rangle}{\sigma_r}.
\end{equation*}
If $|Z|$ $>$ 1.96, then the segregation of our network is significantly different from random. For the variables that we found significant segregation, we  examined the distributions of that attribute across the clusters to better understand the nature of the segregation (i.e., which categories of the attribute were more likely to be assigned to certain clusters).

\subsubsection{Comparing student learning across instruction types}

To address the third research question, we first calculated an individual effect size---a standardized mean difference between students' pre- and post-semester concept inventory scores---for each individual course using Hedges' \textit{g}~\cite{gurevitch1999statistical}. Hedges' \textit{g} is a form of Cohen's \textit{d}~\cite{cohen1992quantitative} that accounts for possible biases due to small sample sizes (e.g., small-enrollment physics courses). Prior studies recommend using Hedges' \textit{g} for all courses in a study even if only a few of them have small sample sizes~\cite{turner2006calculating}.

As in meta-analytic studies~\cite{freeman2014active}, we then combined the individual effect sizes to calculate a net effect size for each cluster (i.e., instruction type). We used a random effects model, which assumes the presence of both within-cluster and between-cluster variation. The weight of each individual effect size, \textit{i}, is calculated as:
\begin{equation*}
    w_i = \frac{1}{v_i + \tau^2}
\end{equation*}
where $v_i$ is the within-cluster variance and $\tau^2$ is the between-cluster variance. The net effect size for each cluster is calculated by taking the sum of each individual effect size multiplied by its weight and dividing that sum by the sum of all the weights:
\begin{equation*}
    g_{net} = \frac{\sum_{i=1}^{k} w_i g_i}{\sum_{i=1}^{k} w_i}
\end{equation*}
where \textit{k} is the number of courses in the cluster and $g_i$ is the individual effect size for course \textit{i}. The 95\% confidence interval of each net effect size is given by $g_{net} \pm (1.96 \times s_{g_{net}})$, where $s_{g_{net}}$ is the standard deviation of the net effect size:
\begin{equation*}
s_{g_{net}} = \sqrt{\frac{1}{\sum_{i=1}^{k} w_i}}.
\end{equation*}

We determined whether the net effect sizes significantly differ between the clusters by conducting a test of heterogeneity. This test quantifies whether the variation between the net effect sizes is greater than chance (i.e., random noise). We calculated the $Q$ statistic:
\begin{equation*}
    Q = \sum_{j=1}^{m} w_j \times (g_{net, j} - \bar g_{net})^2
\end{equation*}
where $m$ is the number of clusters, $w_j$ is the sum of weights within cluster \textit{j}, $g_{net, j}$ is the net effect size for cluster \textit{j}, and 
$\bar g_{net}$ is the mean net effect size across all clusters. The degrees of freedom of $Q$ is equal to one less than the number of clusters. The statistical significance of $Q$, when compared to a chi-squared distribution, indicates whether variation in net effect sizes is due to random chance alone (if $Q$ is not significant) or instead due to some other source of variation (e.g., different types of active learning instruction, if $Q$ is significant).

\section{Results}

In this section, we present the results by research question: the typology of instructional practices used in named active learning methods, the relationships between instruction type and course and institution characteristics, and the relationship between instruction type and student conceptual learning.

\begin{figure*}[t]
  \centering
  \subfloat[]
  {\includegraphics[width=3.1in,trim = {2cm 7cm 2cm 7cm}]{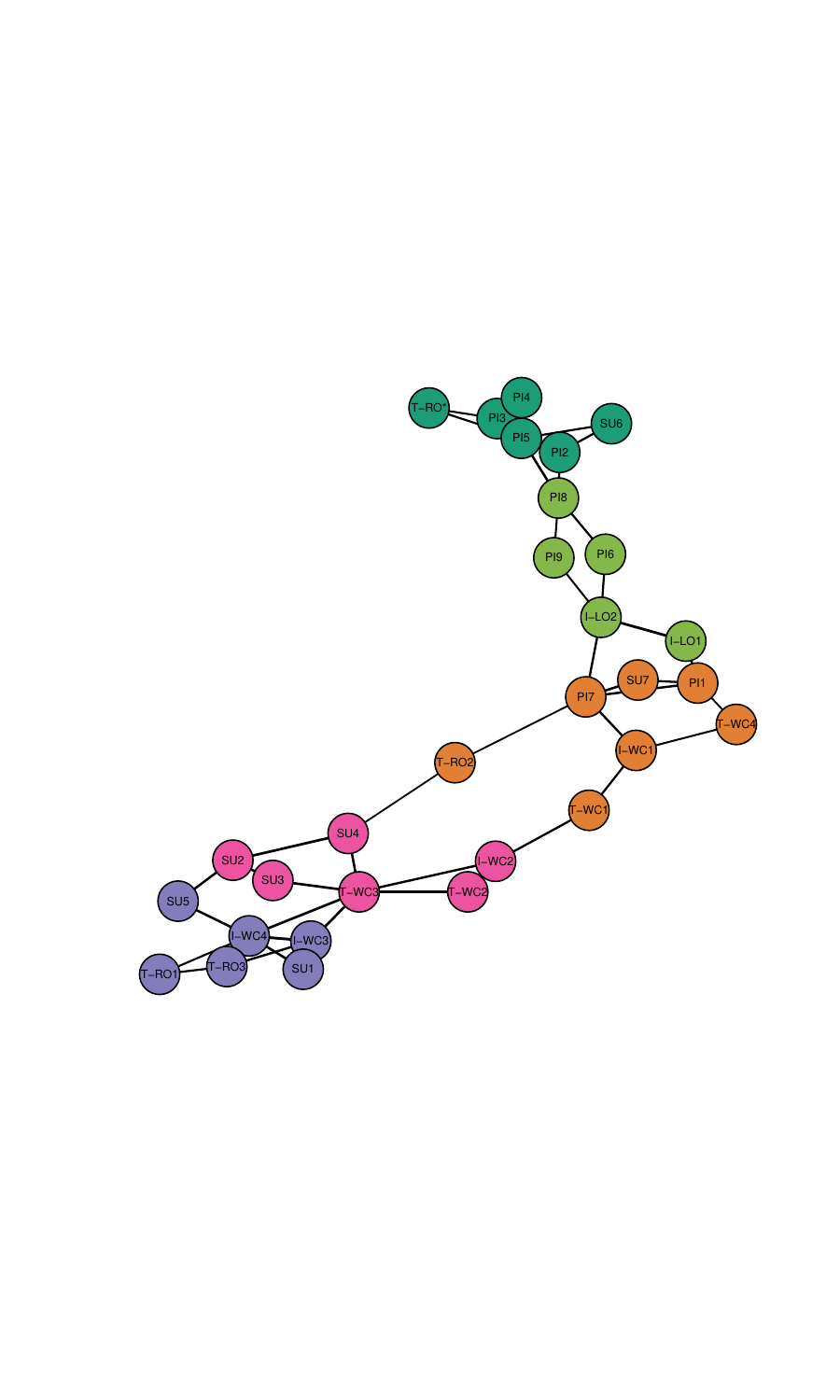}\label{fig:map}}
  \hfill
  \subfloat[]{\includegraphics[width=3.7in]{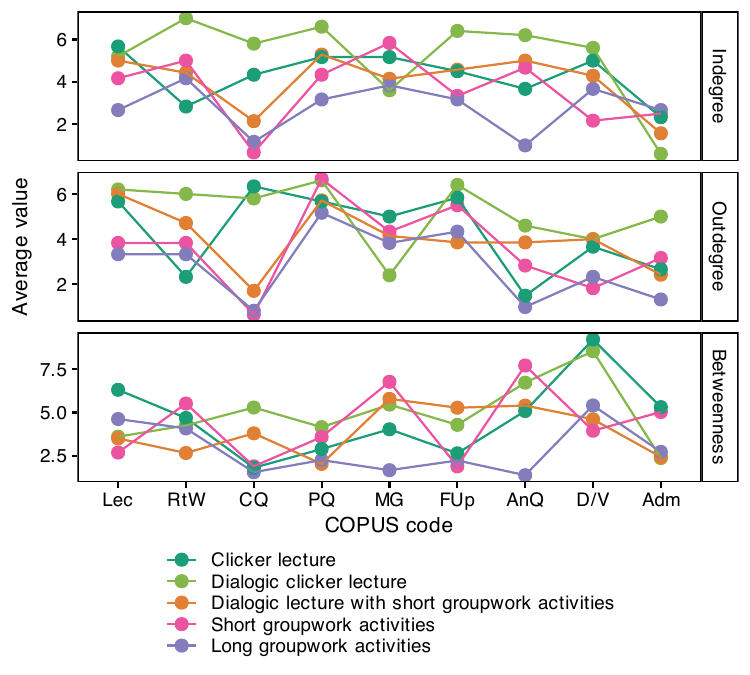}\label{fig:centralities}}
  \caption{(a) Sparsified similarity network with nodes (i.e., classroom observation networks for each instructor) colored by cluster (i.e., instruction type) and edge weights proportional to cosine similarity. I-WC indicates whole-class implementation of ISLE, I-LO indicates lab-only implementation of ISLE (i.e., the observation networks represent lecture sections), PI indicates Peer Instruction, T-WC indicates whole-class implementation of Tutorials, T-RO indicates recitation-only implementation of Tutorials (T-RO* indicates recitation-only implementation of Tutorials but with the observation network representing the lecture section of the course), and SU indicates SCALE-UP. The attributes of each course are provided in Table~\ref{tab:individualfeatures} in the Appendix. (b) Average node-level measures for observation networks assigned to each instruction type (see definitions in Table~\ref{tab:networkdefs}). \label{networks}}
\end{figure*}

\begin{table*}[t] 
\centering
\caption{\label{tab:networkfeatures}
Average network-level measures for classroom observation networks assigned to each instruction type (see definitions in Table~\ref{tab:networkdefs}).}
\begin{ruledtabular}
\setlength{\extrarowheight}{1pt}
\begin{tabular}{lcccc}
Instruction type & Edges & Total strength & Mutuality & Transitivity   \\
\hline
Clicker lecture & 38.7 & 1.02 & 0.72 & 0.83 \\
Dialogic clicker lecture & 47.0 & 1.24 & 0.75 & 0.85 \\
Dialogic lecture with short groupwork activities  & 36.4 & 0.71 & 0.72 & 0.82 \\
Short groupwork activities   & 32.7 & 0.59 & 0.62 & 0.82 \\
Long groupwork activities & 25.5 & 0.47 & 0.52 & 0.64 \\
\end{tabular} 
\end{ruledtabular}
\end{table*}

\subsection{Typology of active learning}

The Infomap community detection algorithm identifies five clusters in the sparsified similarity network, each containing five to seven classroom observation networks (i.e., instructors; Fig.~\ref{fig:map}). Below we use the average node-level and network-level measures (Fig.~\ref{fig:centralities} and Table~\ref{tab:networkfeatures}) for classroom observation networks assigned to each cluster to characterize the type of instruction the cluster represents. These five instruction types comprise our typology of active learning.

\begin{figure*}[t]
  \centering
  \subfloat[Clicker lecture (Peer Instruction, PI3)]
  {\includegraphics[width=2.3in,trim = {2cm 3cm 2cm 2cm}]{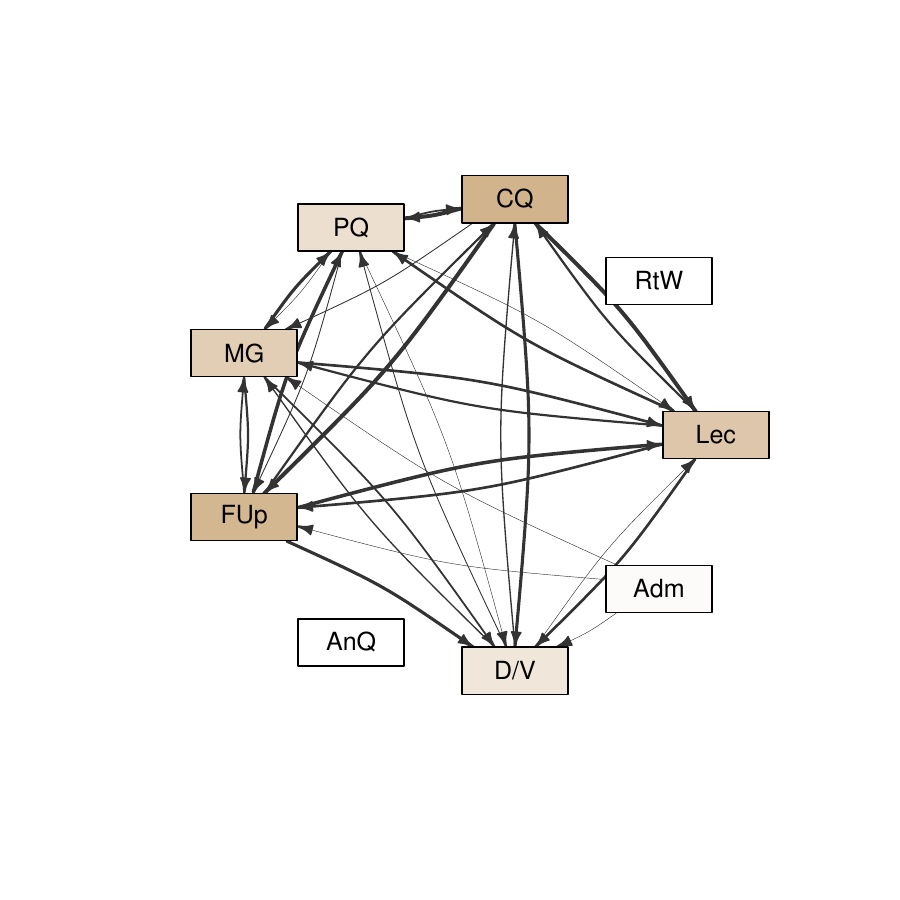}\label{fig:clust1}}
  \hfill
   \subfloat[Dialogic clicker lecture (ISLE lab-only implementation, I-LO2)]
  {\includegraphics[width=2.3in,trim = {2cm 3cm 2cm 2cm}]{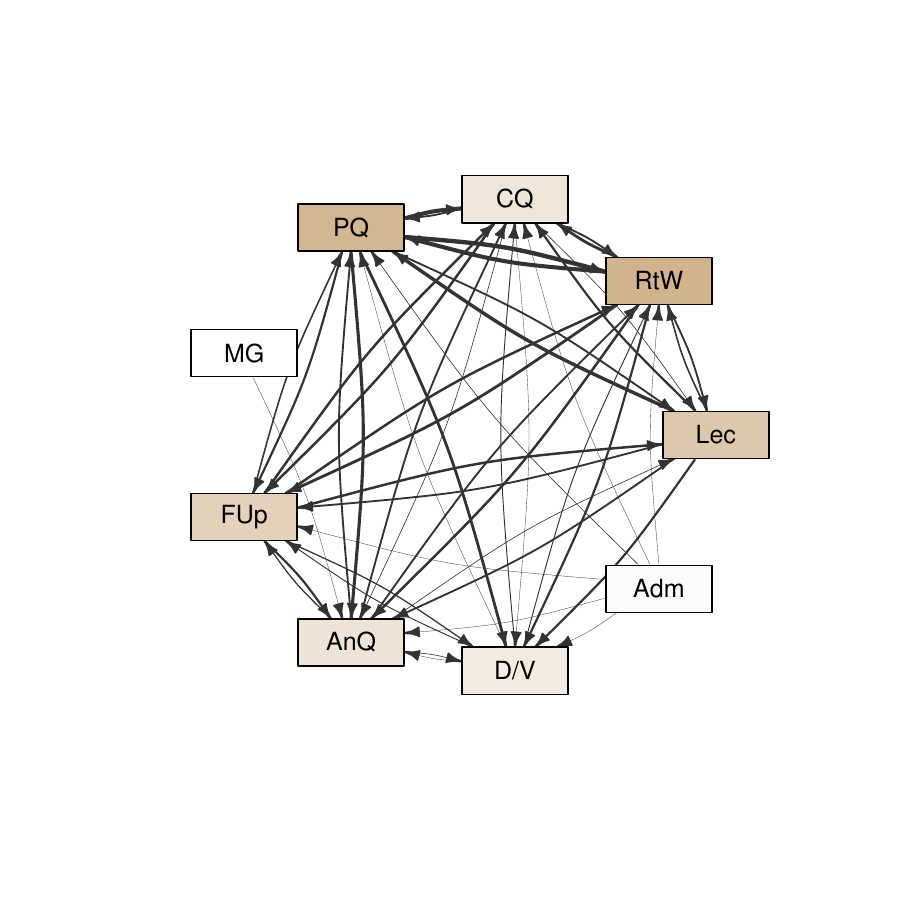}\label{fig:clust2}}
  \hfill
   \subfloat[Dialogic lecture with short groupwork activities (Tutorials whole-class implementation, T-WC1)]
  {\includegraphics[width=2.3in,trim = {2cm 3cm 2cm 2cm}]{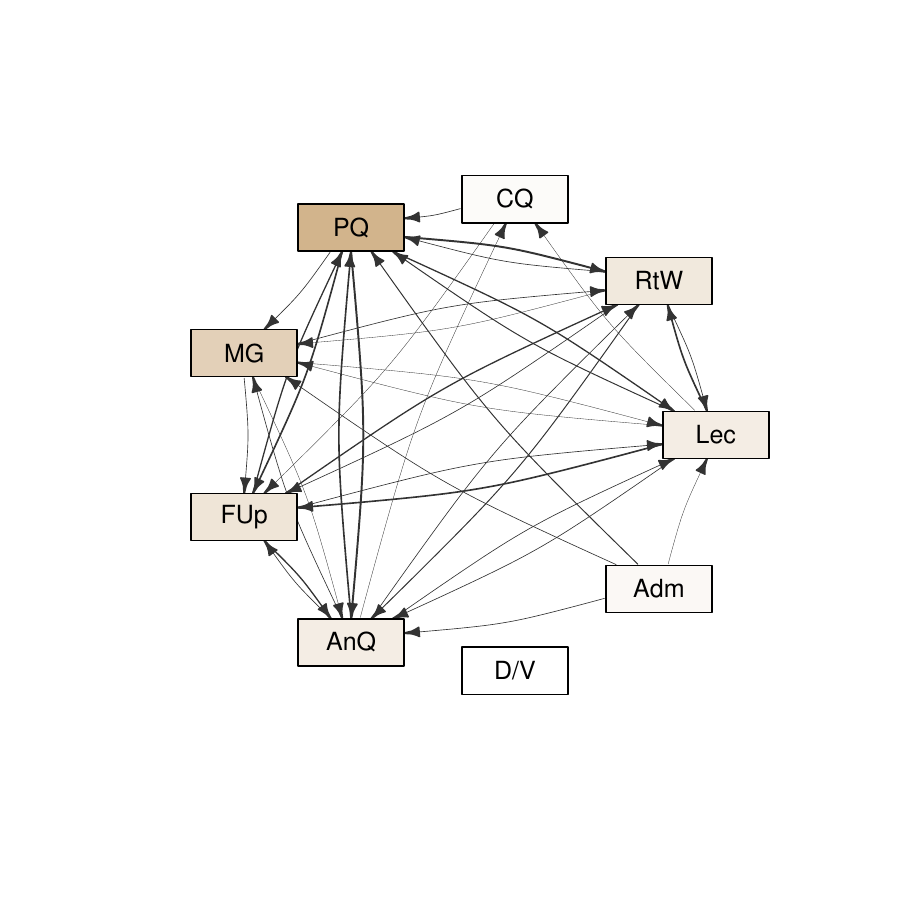}\label{fig:clust3}}
  \hfill
   \subfloat[Short groupwork activities (SCALE-UP, SU3)]
  {\includegraphics[width=2.3in,trim = {2cm 3cm 2cm 2cm}]{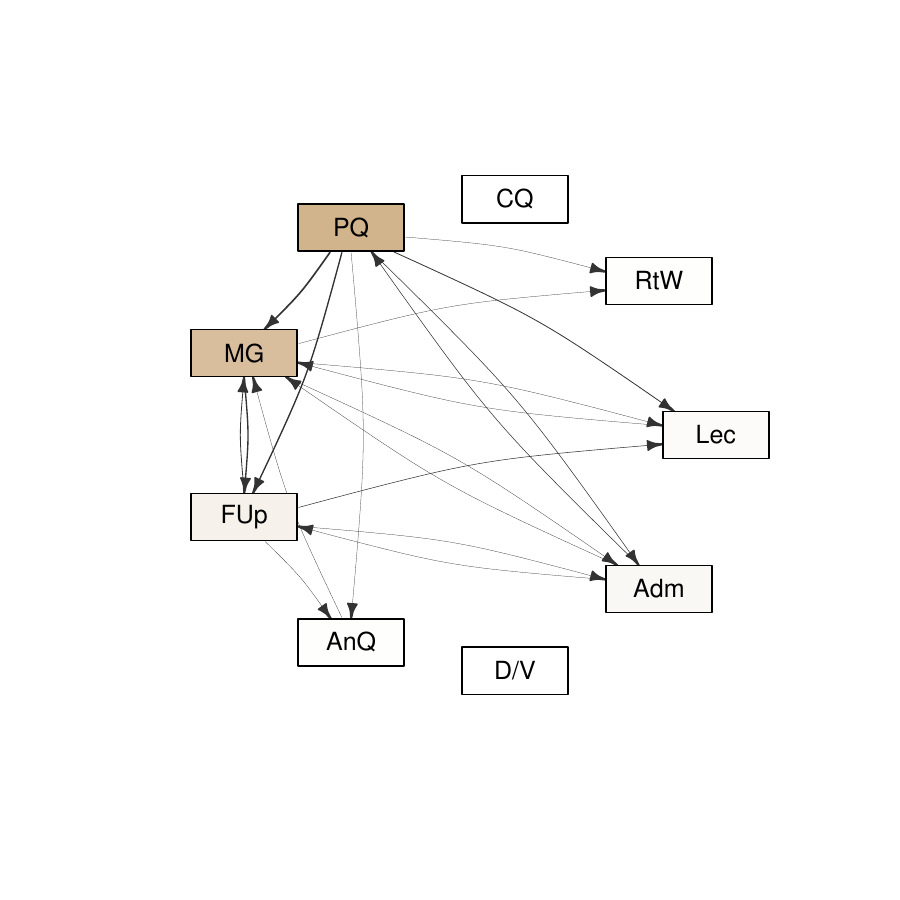}\label{fig:clust4}}
  %\hfill
   \subfloat[Long groupwork activities (Tutorials recitation-only implementation, T-RO1)]
  {\includegraphics[width=2.3in,trim = {2cm 3cm 2cm 2cm}]{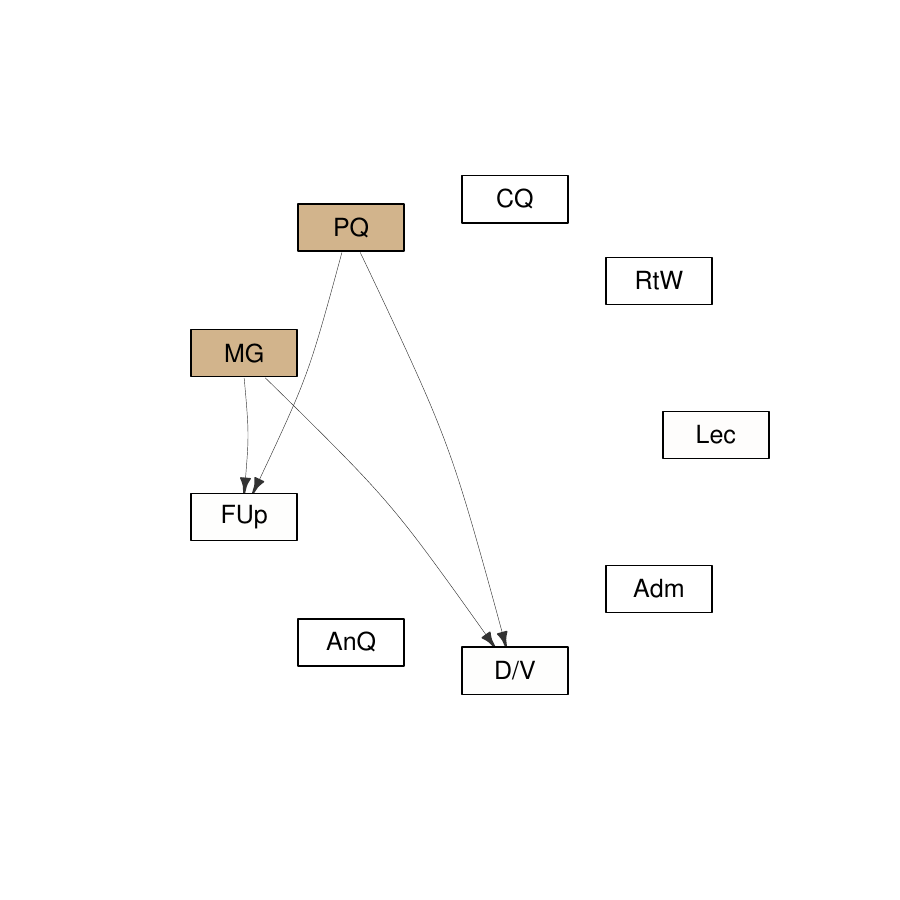}\label{fig:clust5}}
  \caption{Example classroom observation networks for each  instruction type. The active learning method and course label (as in Fig.~\ref{fig:map} and Table~\ref{tab:individualfeatures} in the Appendix) for each network is indicated in parentheses in the sub-captions. Though not directly part of our analysis, node color indicates the fraction of all two-minute time intervals that the code was present: darker colors indicate larger fractions (i.e., longer time durations). Edges point from an initial code to the code that occurs in the following two-minute time interval of the COPUS observation. Edge width indicates the number of transitions occurring between those two codes normalized by the total number of two-minute time intervals observed. All 30 classroom observation networks are available at Ref.~\cite{github2025}.\label{examplenetworks}}
\end{figure*}

\subsubsection{Clicker lecture}

The first type of active learning instruction---clicker lecture---is comprised of four Peer Instruction courses, one SCALE-UP course, and one Tutorials course (recitation-only implementation, but with the classroom observation coming from the lecture section; see dark green nodes in Fig.~\ref{fig:map}). Instructors whose classroom observation networks are assigned to this instruction type frequently cycle through lecturing, asking and guiding groups through a clicker question, and following up on that clicker question with a demonstration or video (relatively high indegree and outdegree for all of these codes and high betweenness for lecturing and demonstrations/videos, see dark green points in Fig.~\ref{fig:centralities}; relatively high edges, total strength, mutuality, and transitivity, signifying a more ``complex" instructional style that involves many transitions between different activities, see Table~\ref{tab:networkfeatures}; see Fig.~\ref{fig:clust1} for an example classroom observation network of this instruction type). In this instruction type, follow-up discussions of clicker questions tend to involve quick shout-out responses from all students to questions posed by the instructor (e.g., ``So the answer is?" ``C!"; relatively high indegree and outdegree for posing questions, see dark green points in Fig.~\ref{fig:centralities}).

\subsubsection{Dialogic clicker lecture}

Three Peer Instruction courses and two ISLE courses (both lab-only implementations, recall that these instructors recorded the lecture sections) are assigned to the second instruction type: dialogic clicker lecture (see light green nodes in Fig.~\ref{fig:map}). Similar to the first instruction type, instructors whose classroom observation networks are assigned to this instruction type frequently cycle through lecturing, asking a clicker question, and following up on that clicker question with a demonstration or video and/or real-time writing (relatively high indegree and outdegree for all of these codes and high betweenness for clicker questions, posing questions, following up, and demonstrations/videos; see light green points in Fig.~\ref{fig:centralities}; highest edges, total strength, mutuality, and transitivity of all five instruction types, indicating the most transitions between activities, see Table~\ref{tab:networkfeatures}; see Fig.~\ref{fig:clust2} for an example classroom observation network of this instruction type). Unlike the first instruction type, the lecture and follow-up discussions involve dialogue between the instructor and individual students: the instructor poses questions to which individual students raise their hands and respond and/or individual students ask questions to which the instructor responds (high indegree, outdegree, and betweenness for posing questions and answering questions; see light green points in Fig.~\ref{fig:centralities}).

\begin{figure*}[t]
  \centering
  \subfloat[]
  {\includegraphics[width=2.95in]{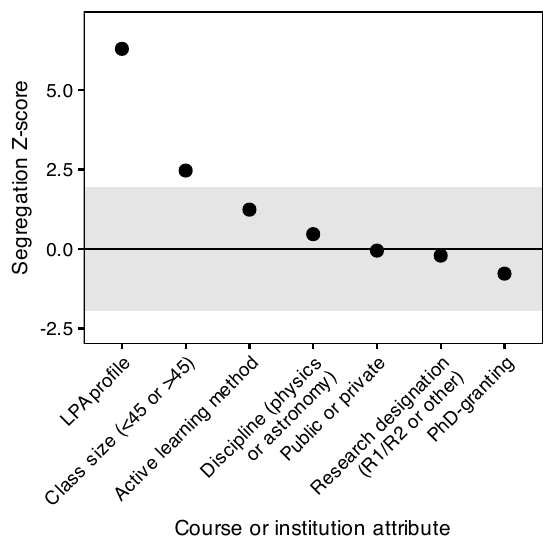}\label{fig:seg}}
  \hfill
  \subfloat[]{\includegraphics[width=4in]{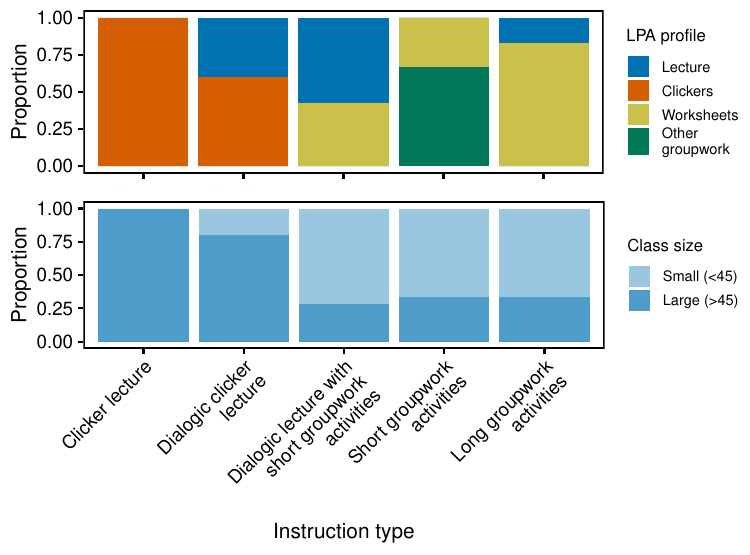}\label{fig:segdist}}
  \caption{(a) Segregation Z-scores for various course and institution attributes. Points indicate the Z-scores and the gray shaded area indicates where the Z-score is not statistically significant, $|Z| < 1.96$. Points falling outside of this region indicate that there is a significant relationship between the corresponding attribute and instruction type. (b) For the two attributes with significant segregation Z-scores, LPA profile and class size, the distributions of courses with these attributes across instruction types. \label{Segregation}}
\end{figure*}

\subsubsection{Dialogic lecture with short groupwork activities}

%Short groupwork activities with dialogic follow-up

The third instruction type---dialogic lecture with short groupwork activities---is the only type that contains classroom observation networks from all four active learning methods: two Peer Instruction courses, one SCALE-UP course, one ISLE course (whole-class implementation), and three Tutorials courses (two whole-class implementations and one recitation-only implementation, see orange nodes in Fig.~\ref{fig:map}). Instructors of this type cycle through lecturing, assigning and guiding various short groupwork activities (some of which are clicker questions as in the first two instruction types, and some of which are other activities such as worksheets or solving problems on whiteboards), and following up on those activities with demonstrations or videos (relatively high indegree and outdegree for lecturing, posing questions, and demonstrations/videos and relatively high betweenness for clicker questions, moving and guiding groupwork, and following up, see orange points in Fig.~\ref{fig:centralities}; moderate number of edges, total strength, mutuality, and transitivity, indicating a moderate amount of transitions between activities, see Table~\ref{tab:networkfeatures}; see Fig.~\ref{fig:clust3} for an example observation network of this instruction type). Similar to the previous instruction type, the lecture and follow-up discussions to groupwork activities are dialogic, with the instructor posing questions to which individual students respond and responding to questions asked by individual students (high indegree and outdegree for posing questions and answering questions, see orange points in Fig.~\ref{fig:centralities}).

\subsubsection{Short groupwork activities}

The next instruction type---short groupwork activities---is comprised of three SCALE-UP courses, two Tutorials courses (both whole-class implementations) and one ISLE course (whole-class implementation, see pink nodes in Fig.~\ref{fig:map}). Instructors of this type do not lecture as much as the previous three instruction types; instead, they cycle through assigning and guiding short groupwork activities (not including clicker questions) and following up on those activities with real-time writing (low node-level measures for lecturing and clicker questions, high node-level measures for moving and guiding groupwork, high outdegree for following up, and high indegree and betweenness for real-time writing, see pink points in Fig.~\ref{fig:centralities}; lower network-level measures than the previous instruction types, see Table~\ref{tab:networkfeatures}; see Fig.~\ref{fig:clust4} for an example observation network of this instruction type). Similar to the previous two instruction types, the follow-up discussions to groupwork activities look like class-wide discussions, with the instructor posing and responding to questions (highest outdegree for posing questions and highest betweenness for answering questions, see pink points in Fig.~\ref{fig:centralities}). Unlike the first three instruction types, this type does not use many demonstrations or videos (relatively low node-level measures for demonstrations/videos, see pink points in Fig.~\ref{fig:centralities}).

\subsubsection{Long groupwork activities}

The final instruction type---long groupwork activities---contains two SCALE-UP courses, two ISLE courses (both whole-class implementations) and two Tutorials course (both recitation-only implementations, see purple nodes in Fig.~\ref{fig:map}). Instructors whose classroom observations are assigned to this instruction type predominantly guide one or a few long groupwork activities (e.g., worksheets) and then follow up on them using real-time writing and/or demonstrations or videos (node-level measures are low compared to the other instruction types for most COPUS codes, but comparable indegree and outdegree to other instruction types for moving and guiding groupwork and following up and moderately high betweenness for real-time writing and demonstrations/videos, see purple points in Fig.~\ref{fig:centralities}; lowest number of edges, total strength, mutuality, and transitivity of all five instruction types, indicating fewer transitions between activities and longer activities, see Table~\ref{tab:networkfeatures}; see Fig.~\ref{fig:clust5} for an example classroom observation network of this instruction type). Follow-ups to the groupwork activities are more didactic than dialogic (low node-level measures for posing questions and answering questions, see purple points in Fig.~\ref{fig:centralities}).

\subsection{Relating instruction types to course and institution attributes}

There is a significant relationship between instruction type and latent profile identified in our prior work ($Z$ = 6.35)~\cite{sundstrom2025relativebenefits} and between instruction type and class size (whether smaller or larger than 45 enrolled students, $Z =$ 2.44; see points falling above the gray highlighted area in Fig.~\ref{fig:seg}). For the latent profiles, the first two instruction types (both clicker lectures) contain all of the courses that were assigned to the clickers profile in our previous analysis (Fig.~\ref{fig:segdist}). The dialogic clicker lecture type also contains several of the courses assigned to the lecture profile in our LPA. Both the dialogic lecture with short groupwork activities and long groupwork activities types contain courses that were assigned to the lecture and worksheets profiles in our previous work. The short groupwork activities type contains several courses assigned to the worksheets profile and all of the courses assigned to the other groupwork profile. For class size, the first two instruction types, clicker lecture and dialogic clicker lecture, contain most of the large classes while the remaining three instruction types contain most of the small classes (Fig.~\ref{fig:segdist}).

Instruction type is not significantly related to active learning method (whether ISLE, Peer Instruction, Tutorials, or SCALE-UP; $Z =$ 1.23), discipline (whether physics or astronomy; $Z =$ 0.47), whether the institution is public or private ($Z =$ --0.04), institution research designation (whether R1/R2 or not according to the 2025 Carnegie Research Classification; $Z =$ --0.23), or PhD-granting status ($Z =$  --0.75, see points falling within the gray shaded area, $|Z| < 1.96$, in Fig.~\ref{fig:seg}). In other words, none of these course or institution attributes correlate with the instruction type used by instructors; instead, courses with the same attribute are spread across the instruction types fairly randomly. With regard to active learning method, this means that there is a large amount of variation in the implementation of the four named active learning methods we study here, as courses from each method are mixed across different instruction types. Named active learning methods, therefore, do not necessarily map onto specific instructional practices: instructors likely adapt the methods to their local context (e.g., student population and available classrooms).

\begin{figure}[t]
    \centering
    \includegraphics[width=3.4in,trim={0 0.3cm 0 0}]{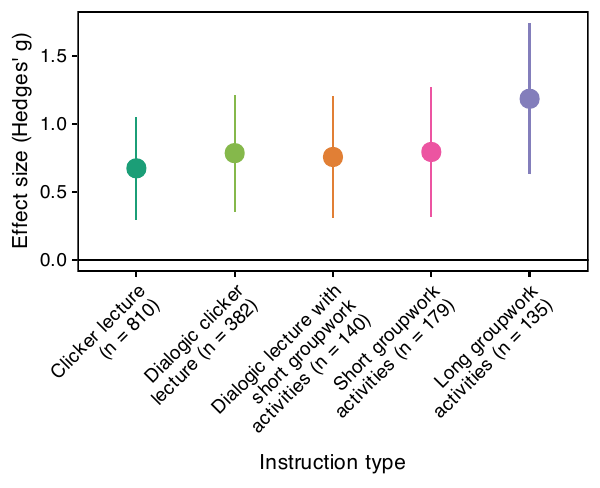}
    \caption{Effect sizes for concept inventory scores by instruction type. Points represent Hedges' \textit{g} values and error bars indicate 95\% confidence intervals. Positive (negative) values indicate increases (decreases) in student scores from pre- to post-semester. \textit{N} values indicate the number of students included in the analysis for each instruction type (i.e., the number of students with matched pre- and post-semester concept inventory scores).
    }
    \label{fig:Learning}
\end{figure}

\subsection{Relating instruction types to student conceptual learning}

A heterogeneity analysis indicates no statistically significant variation in student conceptual learning gains based on instruction type ($Q =$ 2.32, \textit{df} $=$ 4, \textit{p} $=$ 0.68). Qualitatively, the effect size for the last instruction type---long groupwork activities---is slightly larger than the other four instruction types; however, the confidence intervals of Hedges' \textit{g} for all instruction types overlap with one another (Fig.~\ref{fig:Learning}). This means that not only are established active learning methods implemented in different ways (i.e., we saw no significant segregation of active learning methods across instruction types), but also that certain implementations are not necessarily more effective than others for student conceptual learning.

\section{Discussion}

We have employed a novel method for characterizing and comparing instructional practices based on classroom observation networks. Such networks allow for researchers to visualize chronological transitions between classroom activities and directly measure similarity between pairs of observations. Applying this method to a dataset of 30 instructors implementing one of four named active learning methods (ISLE, Peer Instruction, Tutorials, and SCALE-UP) in their introductory physics or astronomy course, we identified five instruction types: clicker lecture, dialogic clicker lecture, dialogic lecture with short groupwork activities, short groupwork activities, and long groupwork activities. 

This typology of active learning instruction not only corroborates our previous latent profile analysis of these data~\cite{sundstrom2025relativebenefits}, providing some validity to both groupings of observations, but also provides a far more detailed framework for describing instructional practices than existing literature. Smith and colleagues, for example, found that ``most faculty members fall somewhere in the continuum between pure lecturing and and primarily active-engagement instruction"~\cite{smith2014campus} (p. 634) based on descriptive statistics of classroom observations. The current study provides a more nuanced characterization of this continuum that includes information about specific sequences of activities used by instructors. 

The analysis presented here also substantiates prior work highlighting that instructors do not adopt existing active learning methods as is; rather, they adapt them to their instructional context (e.g., student population and physical classroom space)~\cite{borrego2013fidelity,dancy2016implement,scanlon2019method}. We found that instruction type is not significantly related to named active learning method. Instead, instructors using each method were distributed across the identified instruction types. This result points to a possible shift in the way we think and talk about active learning: using the names of developed active learning methods may not actually reflect the underlying activities involved in the instruction.

We also noticed substantial variation \textit{within} each instruction type, signaling that a broader typology of active learning instruction (e.g., with more than five types) likely exists. We recommend for future research studies to conduct a similar analysis with a broader range of active learning courses (i.e., not only those using named active learning methods) to build on our typology. Future work may also map instructors' perceived barriers to adopting active learning strategies~\cite{henderson2007barriers,henderson2012use,dancy2012experiences,Affriyenni2025navigating} and/or received training (e.g., whether they have previously attended the Physics and Astronomy
Faculty Teaching Institute) to their instruction types (e.g., by also surveying or interviewing participating instructors). This line of work would offer two important implications. First, researchers could start to unpack some of the mechanisms (i.e., beyond class size) underlying different implementations of the same named active learning method, as observed here. It is plausible, for example, that instructors who received general training about active learning implement a given method differently from instructors who received training about the particular method (e.g., by attending a workshop). Second, in professional development settings, we could provide physics instructors with examples of what different implementations of each method look like and a guide to choosing a particular implementation based on their personal teaching philosophies, perceived barriers at their institutions, and other factors (e.g., similar to Ref.~\cite{scherr2007enabling}).

Surprisingly, variables related to institution type (i.e., public or private, research designation, and PhD-granting or not) were also not significantly related to instruction type despite existing research indicating that level of institutional support is often an important factor for instructors deciding whether or not to adopt active learning strategies~\cite{henderson2007barriers,henderson2012use,dancy2012experiences,Affriyenni2025navigating}. Our measures of institutional support, however, were likely too broad (e.g., representative of the entire institution rather than the physics or astronomy department) to capture any meaningful relationships (if they exist). We suggest for future work to collect more detailed information from the instructor participants about support from their department to make more rigorous claims about the role of institutional support in instructor adaptations to active learning methods.

At the same time, we identified that one of the measured course and institution attributes---class size---does relate to instruction type. Large classes (containing more than 45 students) were more often assigned to one of the two clicker lecture types and small classes were more often assigned to one of the remaining three instruction types. A few of the large classes, however, were assigned to dialogic lecture with short groupwork activities, short groupwork activities, or long groupwork activities. This means that despite documented challenges to incorporating active learning strategies into larger courses~\cite{murdoch2002active}, it is possible to do so \textit{and} in a way that involves class-wide dialogue and/or a range of groupwork activities (i.e., not only clickers). Researchers should conduct controlled experiments to compare the implementations and impacts (e.g., on conceptual learning) of active learning methods in different class sizes.

We also expand upon the few existing studies that relate specific instructional practices to student conceptual learning~\cite{weir2019small,bukola2025perc,sundstrom2025relativebenefits,cleveland2017investigating,connell2016increasing,mcneal2020biosensors}. In our previous analysis of this dataset, we identified that while all four active learning methods exhibited conceptual learning gains larger than zero, SCALE-UP exhibited significantly larger conceptual learning gains than both ISLE and Peer Instruction~\cite{sundstrom2025relativebenefits}. In the current study, however, we found no significant differences in student conceptual learning across the instruction types. Taken together with the other results from the current study, this suggests that not only are named active learning methods modified in practice, but also that instructor modifications do not reduce the effectiveness of the methods for student conceptual learning. In other words, effective active learning likely does not require strict adherence to the method developers' intentions: instructors can be flexible in adapting the methods to their own needs. We suggest for future research to investigate additional explanations for why SCALE-UP may be more effective for student learning than other methods~\cite{sundstrom2025relativebenefits}, as this study suggests that specific classroom activities likely do not provide a complete explanation. Other possible explanations include the quality or fidelity of classroom activities (e.g., the SCALE-UP courses in our dataset may have been higher fidelity implementations of the method than the courses using the other three methods) and the nature of the methods in and of themselves (e.g., SCALE-UP provides one cohesive picture of the course material by blending lecture, problem solving, and lab activities).

We acknowledge several limitations to this work that prompt further research. Related to instructor recruitment, most instructors included in this study are physics education researchers themselves or fairly involved in the physics education research community. Yet research shows that science education researchers employing active learning methods in their classroom teach differently and exhibit more significant improvements to student outcomes than a random sample of instructors employing active learning methods~\cite{andrews2011active}. Future research, therefore, should aim to collect similar data from a broader sample of physics instructors, including information about the instructors' familiarity with physics education research.

In terms of data collection, we asked instructors to record three consecutive class sessions. Instructors interpreted these instructions differently (e.g., some lab-only implementations of ISLE only recorded the lecture sections, where the method was not implemented) and the variety of course structures limited the consistency of the video recordings (e.g., in some---but not all---of the Peer Instruction classes, students attended recitation and laboratory sections in addition to lecture, but these were not recorded). We recommend for future iterations of this work to collect video recordings of one full week of course contact hours, capturing all course components that a student would attend in a given week.

We chose to use the COPUS for our analysis of the classroom observations. The instructor codes in the COPUS distinguish clicker questions from other forms of groupwork, but they do not tease apart the instructor assigning or proctoring other types of groupwork such as worksheets versus laboratory activities. The COPUS was also designed for descriptive, frequency-based analyses and not the temporal analysis conducted here~\cite{smith2013classroom}. Future research, therefore, may add more refined groupwork codes to the COPUS and/or conduct a similar analysis using a different observation protocol that uses continuous (rather than interval-based) coding (e.g., the Framework for Interactive Learning in Lectures, FILL+~\cite{wood2016characterizing,kinnear2021developing}).

Finally, we measured student conceptual learning as one outcome that may be impacted by different types of active learning instruction. We recommend for future work to measure additional outcomes of interest, such as attitudes toward science, critical thinking skills, and sense of belonging, to understand the full scope of effects of instruction type on students.

\section{Conclusion}

We have advanced previous descriptive studies of active learning instruction by capturing the temporal and interactional nature of instructional practices. We hope that the typology of active learning instruction defined here serves as a starting set of instruction types for future studies to build upon. We encourage these future studies to incorporate data from a broader range of active learning strategies (i.e., not only named methods), instructors (i.e., not only those who are involved in science education research), courses (i.e., not only introductory level), scientific disciplines (e.g., biology, chemistry, and engineering), and institutions (e.g., outside of the United States). Further iterations of our typology have the potential to improve the professional development of college physics instructors by providing them with examples of modifications to existing active learning methods that have been used in a variety of instructional contexts.

\section*{ACKNOWLEDGEMENTS}

We thank all of the instructors and students who participated in our study. We also thank Ibukunoluwa Bukola for meaningful feedback on this manuscript. This material is based upon work supported by the National Science Foundation under Grant Nos. 2111128 and 2111275. M. S. is partly funded by the Cotswold Foundation Postdoctoral Fellowship at Drexel University.

\bibliography{COPUSNets}

\begin{thebibliography}{10}

\bibitem{piaget1926language}
Jean Piaget.
\newblock {\em The language and thought of the child}.
\newblock Harcourt Brace, New York, 1926.

\bibitem{vygotsky1978mind}
Lev~S. Vygotsky.
\newblock {\em Mind in society: The development of higher psychological processes}, volume~86.
\newblock Harvard University Press, 1978.

\bibitem{hake1998interactive}
Richard~R. Hake.
\newblock Interactive-engagement versus traditional methods: A six-thousand-student survey of mechanics test data for introductory physics courses.
\newblock {\em American Journal of Physics}, 66(1):64--74, 1998.

\bibitem{freeman2014active}
Scott Freeman, Sarah~L. Eddy, Miles McDonough, Michelle~K. Smith, Nnadozie Okoroafor, Hannah Jordt, and Mary~Pat Wenderoth.
\newblock Active learning increases student performance in science, engineering, and mathematics.
\newblock {\em Proceedings of the National Academy of Sciences}, 111(23):8410--8415, 2014.

\bibitem{theobald2020active}
Elli~J. Theobald, Mariah~J. Hill, Elisa Tran, Sweta Agrawal, E.~Nicole Arroyo, Shawn Behling, Nyasha Chambwe, Dianne~Laboy Cintr{\'o}n, Jacob~D. Cooper, Gideon Dunster, et~al.
\newblock Active learning narrows achievement gaps for underrepresented students in undergraduate science, technology, engineering, and math.
\newblock {\em Proceedings of the National Academy of Sciences}, 117(12):6476--6483, 2020.

\bibitem{mazur1997peer}
Eric Mazur.
\newblock {\em Peer Instruction: A User's Manual}.
\newblock Prentice Hall, 1997.

\bibitem{crouch2007peer}
Catherine~H. Crouch, Jessica Watkins, Adam~P. Fagen, and Eric Mazur.
\newblock Peer instruction: Engaging students one-on-one, all at once.
\newblock {\em Research-based Reform of University Physics}, 1(1):40--95, 2007.

\bibitem{etkina2007investigative}
Eugenia Etkina and Alan Van~Heuvelen.
\newblock Investigative science learning environment--a science process approach to learning physics.
\newblock {\em Research-based Reform of University Physics}, 1(1):1--48, 2007.

\bibitem{etkina2015millikan}
Eugenia Etkina.
\newblock Millikan award lecture: Students of physics—listeners, observers, or collaborative participants in physics scientific practices?
\newblock {\em American Journal of Physics}, 83(8):669--679, 2015.

\bibitem{henderson2007barriers}
Charles Henderson and Melissa~H. Dancy.
\newblock Barriers to the use of research-based instructional strategies: The influence of both individual and situational characteristics.
\newblock {\em Physical Review Special Topics—Physics Education Research}, 3(2):020102, 2007.

\bibitem{henderson2012use}
Charles Henderson, Melissa Dancy, and Magdalena Niewiadomska-Bugaj.
\newblock Use of research-based instructional strategies in introductory physics: Where do faculty leave the innovation-decision process?
\newblock {\em Physical Review Special Topics—Physics Education Research}, 8(2):020104, 2012.

\bibitem{dancy2012experiences}
Melissa~H. Dancy and Charles Henderson.
\newblock Experiences of new faculty implementing research-based instructional strategies.
\newblock In {\em AIP Conference Proceedings}, volume 1413, pages 163--166. American Institute of Physics, 2012.

\bibitem{Affriyenni2025navigating}
Yessi Affriyenni, Helen Georgiou, and Noah Finkelstein.
\newblock Navigating the adoption of research-based instructional strategies within the complex nature of higher education.
\newblock {\em Physical Review Physics Education Research}, 21:020124, Sep 2025.

\bibitem{foote2014diffusion}
Kathleen~T. Foote, Xaver Neumeyer, Charles Henderson, Melissa~H. Dancy, and Robert~J. Beichner.
\newblock Diffusion of research-based instructional strategies: The case of {SCALE-UP}.
\newblock {\em International Journal of STEM Education}, 1(1):10, 2014.

\bibitem{henderson2009impact}
Charles Henderson and Melissa~H. Dancy.
\newblock Impact of physics education research on the teaching of introductory quantitative physics in the united states.
\newblock {\em Physical Review Special Topics—Physics Education Research}, 5(2):020107, 2009.

\bibitem{dancy2024physics}
Melissa Dancy, Charles Henderson, Naneh Apkarian, Estrella Johnson, Marilyne Stains, Jeffrey~R. Raker, and Alexandra Lau.
\newblock Physics instructors’ knowledge and use of active learning has increased over the last decade but most still lecture too much.
\newblock {\em Physical Review Physics Education Research}, 20(1):010119, 2024.

\bibitem{borrego2013fidelity}
Maura Borrego, Stephanie Cutler, Michael Prince, Charles Henderson, and Jeffrey~E. Froyd.
\newblock Fidelity of implementation of research-based instructional strategies {(RBIS)} in engineering science courses.
\newblock {\em Journal of Engineering Education}, 102(3):394--425, 2013.

\bibitem{dancy2016implement}
Melissa Dancy, Charles Henderson, and Chandra Turpen.
\newblock How faculty learn about and implement research-based instructional strategies: The case of {Peer Instruction}.
\newblock {\em Physical Review Physics Education Research}, 12:010110, Feb 2016.

\bibitem{scanlon2019method}
Erin Scanlon, Brian Zamarripa~Roman, Elijah Ibadlit, and Jacquelyn~J. Chini.
\newblock A method for analyzing instructors’ purposeful modifications to research-based instructional strategies.
\newblock {\em International Journal of STEM Education}, 6(1):12, 2019.

\bibitem{andrews2011active}
Tessa~M. Andrews, Michael~J. Leonard, Clinton~A. Colgrove, and Steven~T. Kalinowski.
\newblock Active learning not associated with student learning in a random sample of college biology courses.
\newblock {\em CBE-Life Sciences Education}, 10(4):394--405, 2011.

\bibitem{ebert2011we}
Diane Ebert-May, Terry~L. Derting, Janet Hodder, Jennifer~L. Momsen, Tammy~M. Long, and Sarah~E. Jardeleza.
\newblock What we say is not what we do: Effective evaluation of faculty professional development programs.
\newblock {\em BioScience}, 61(7):550--558, 2011.

\bibitem{turpen2009not}
Chandra Turpen and Noah~D. Finkelstein.
\newblock Not all interactive engagement is the same: Variations in physics professors’ implementation of peer instruction.
\newblock {\em Physical Review Special Topics—Physics Education Research}, 5(2):020101, 2009.

\bibitem{smith2014campus}
Michelle~K. Smith, Erin~L. Vinson, Jeremy~A. Smith, Justin~D. Lewin, and MacKenzie~R. Stetzer.
\newblock A campus-wide study of {STEM} courses: New perspectives on teaching practices and perceptions.
\newblock {\em CBE—Life Sciences Education}, 13(4):624--635, 2014.

\bibitem{wood2016characterizing}
Anna~K. Wood, Ross~K. Galloway, Robyn Donnelly, and Judy Hardy.
\newblock Characterizing interactive engagement activities in a flipped introductory physics class.
\newblock {\em Physical Review Physics Education Research}, 12(1):010140, 2016.

\bibitem{commeford2021characterizing}
Kelley Commeford, Eric Brewe, and Adrienne Traxler.
\newblock Characterizing active learning environments in physics using network analysis and classroom observations.
\newblock {\em Physical Review Physics Education Research}, 17(2):020136, 2021.

\bibitem{bukola2025perc}
Ibukunoluwa Bukola, Meagan Sundstrom, Justin Gambrell, Olive Ross, Adrienne~L. Traxler, and Eric Brewe.
\newblock Multi-institutional assessment of peer instruction implementation and impacts using the framework for interactive learning in lectures.
\newblock {\em arXiv preprint arXiv:2508.08422}, 2025.

\bibitem{weir2019small}
Laura~K. Weir, Megan~K. Barker, Lisa~M. McDonnell, Natalie~G. Schimpf, Tamara~M. Rodela, and Patricia~M. Schulte.
\newblock Small changes, big gains: A curriculum-wide study of teaching practices and student learning in undergraduate biology.
\newblock {\em PLoS One}, 14(8):e0220900, 2019.

\bibitem{sundstrom2025relativebenefits}
Meagan Sundstrom, Justin Gambrell, Colin Green, Adrienne~L. Traxler, and Eric Brewe.
\newblock Relative benefits of different active learning methods to conceptual physics learning.
\newblock {\em arXiv preprint arXiv:2505.04577}, 2025.

\bibitem{cleveland2017investigating}
Lacy~M. Cleveland, Jeffrey~T. Olimpo, and Sue~Ellen DeChenne-Peters.
\newblock Investigating the relationship between instructors’ use of active-learning strategies and students’ conceptual understanding and affective changes in introductory biology: A comparison of two active-learning environments.
\newblock {\em CBE—Life Sciences Education}, 16(2):ar19, 2017.

\bibitem{connell2016increasing}
Georgianne~L. Connell, Deborah~A. Donovan, and Timothy~G. Chambers.
\newblock Increasing the use of student-centered pedagogies from moderate to high improves student learning and attitudes about biology.
\newblock {\em CBE—Life Sciences Education}, 15(1):ar3, 2016.

\bibitem{mcneal2020biosensors}
Karen~S. McNeal, Min Zhong, Nick~A. Soltis, Lindsay Doukopoulos, Elijah~T. Johnson, Stephanie Courtney, Akilah Alwan, and Mallory Porch.
\newblock Biosensors show promise as a measure of student engagement in a large introductory biology course.
\newblock {\em CBE—Life Sciences Education}, 19(4):ar50, 2020.

\bibitem{anwar2021systematic}
Saira Anwar and Muhsin Menekse.
\newblock A systematic review of observation protocols used in postsecondary {STEM} classrooms.
\newblock {\em Review of Education}, 9(1):81--120, 2021.

\bibitem{sawada2002measuring}
Daiyo Sawada, Michael~D. Piburn, Eugene Judson, Jeff Turley, Kathleen Falconer, Russell Benford, and Irene Bloom.
\newblock Measuring reform practices in science and mathematics classrooms: The reformed teaching observation protocol.
\newblock {\em School Science and Mathematics}, 102(6):245--253, 2002.

\bibitem{wainwright2003development}
Camille~L. Wainwright, Lawrence Flick, and Patricia Morrell.
\newblock The development of instruments for assessment of instructional practices in standards-based teaching.
\newblock {\em Journal of Mathematics and Science: Collaborative Explorations}, 6(1):21--46, 2003.

\bibitem{harris2003developing}
Alene~H. Harris and Monica~Farmer Cox.
\newblock Developing an observation system to capture instructional differences in engineering classrooms.
\newblock {\em Journal of Engineering Education}, 92(4):329--336, 2003.

\bibitem{kern2007cooperative}
Anne~L. Kern, Tamara~J. Moore, and F.~Caglin Akillioglu.
\newblock Cooperative learning: Developing an observation instrument for student interactions.
\newblock In {\em 37th annual Frontiers in Education Conference--Global Engineering: Knowledge without Borders, Opportunities without Passports}, pages T1D--1. IEEE, 2007.

\bibitem{hora2013instructional}
Matthew~Tadashi Hora and Joseph~J. Ferrare.
\newblock Instructional systems of practice: A multidimensional analysis of math and science undergraduate course planning and classroom teaching.
\newblock {\em Journal of the Learning Sciences}, 22(2):212--257, 2013.

\bibitem{hora2015toward}
Matthew~T. Hora.
\newblock Toward a descriptive science of teaching: How the {TDOP} illuminates the multidimensional nature of active learning in postsecondary classrooms.
\newblock {\em Science Education}, 99(5):783--818, 2015.

\bibitem{smith2013classroom}
Michelle~K. Smith, Francis H.~M. Jones, Sarah~L. Gilbert, and Carl~E. Wieman.
\newblock The classroom observation protocol for undergraduate {STEM} ({COPUS}): A new instrument to characterize university {STEM} classroom practices.
\newblock {\em CBE—Life Sciences Education}, 12(4):618--627, 2013.

\bibitem{kothiyal2013effect}
Aditi Kothiyal, Rwitajit Majumdar, Sahana Murthy, and Sridhar Iyer.
\newblock Effect of think-pair-share in a large {CS1} class: 83\% sustained engagement.
\newblock In {\em Proceedings of the Ninth Annual International ACM Conference on International Computing Education Research}, pages 137--144, 2013.

\bibitem{finelli2014classroom}
Cynthia~J. Finelli, Matthew DeMonbron, Prateek Shekhar, Maura Borrego, Charles Henderson, Michael Prince, and Cindy~K. Waters.
\newblock A classroom observation instrument to assess student response to active learning.
\newblock In {\em 2014 IEEE Frontiers in Education Conference (FIE) Proceedings}, pages 1--4. IEEE, 2014.

\bibitem{lund2015best}
Travis~J. Lund, Matthew Pilarz, Jonathan~B. Velasco, Devasmita Chakraverty, Kaitlyn Rosploch, Molly Undersander, and Marilyne Stains.
\newblock The best of both worlds: Building on the {COPUS and RTOP} observation protocols to easily and reliably measure various levels of reformed instructional practice.
\newblock {\em CBE—Life Sciences Education}, 14(2):ar18, 2015.

\bibitem{denaro2021comparison}
Kameryn Denaro, Brian Sato, Ashley Harlow, Andrea Aebersold, and Mayank Verma.
\newblock Comparison of cluster analysis methodologies for characterization of {Classroom Observation Protocol for Undergraduate STEM (COPUS)} data.
\newblock {\em CBE—Life Sciences Education}, 20(1):ar3, 2021.

\bibitem{wan2020characterizing}
Tong Wan, Ashley~A. Geraets, Constance~M. Doty, Erin K.~H. Saitta, and Jacquelyn~J. Chini.
\newblock Characterizing science graduate teaching assistants’ instructional practices in reformed laboratories and tutorials.
\newblock {\em International Journal of STEM Education}, 7(1):30, 2020.

\bibitem{campbell2017comprehensive}
Corbin~M. Campbell, Alberto~F. Cabrera, Jessica Ostrow~Michel, and Shikha Patel.
\newblock From comprehensive to singular: A latent class analysis of college teaching practices.
\newblock {\em Research in Higher Education}, 58(6):581--604, 2017.

\bibitem{stains2018anatomy}
Marilyne Stains, Jordan Harshman, Megan~K. Barker, Stephanie~V. Chasteen, Renee Cole, Sue~Ellen DeChenne-Peters, M.~Kevin Eagan~Jr, Joan~M. Esson, Jennifer~K. Knight, Frank~A. Laski, et~al.
\newblock Anatomy of {STEM} teaching in {North American} universities.
\newblock {\em Science}, 359(6383):1468--1470, 2018.

\bibitem{commeford2022characterizing}
Kelley Commeford, Eric Brewe, and Adrienne Traxler.
\newblock Characterizing active learning environments in physics using latent profile analysis.
\newblock {\em Physical Review Physics Education Research}, 18(1):010113, 2022.

\bibitem{weston2023measures}
Timothy~J. Weston, Sandra~L. Laursen, and Charles~N. Hayward.
\newblock Measures of success: Characterizing teaching and teaching change with segmented and holistic observation data.
\newblock {\em International Journal of STEM Education}, 10(1):24, 2023.

\bibitem{stains2019genome}
Robert~M. Erdmann and Marilyne Stains.
\newblock Classroom as genome: Using the tools of genomics and bioinformatics to illuminate classroom observation data.
\newblock {\em CBE—Life Sciences Education}, 18(1):es1, 2019.

\bibitem{hora2014remeasuring}
Matthew~T. Hora and Joseph~J. Ferrare.
\newblock Remeasuring postsecondary teaching: How singular categories of instruction obscure the multiple dimensions of classroom practice.
\newblock {\em Journal of College Science Teaching}, 43(3):36--41, 2014.

\bibitem{bruun2025network}
Jesper Bruun, Linda Udby, and Pia~Jensen Ray.
\newblock Network analyses of student interactions with online textbook problems.
\newblock {\em European Journal of Physics}, 46(2):025704, 2025.

\bibitem{github2025}
\url{https://github.com/msundstrom33/ComparingActiveLearningMethods.git}.

\bibitem{mcdermott2002tutorials}
Lillian~C. McDermott.
\newblock {\em Tutorials in Introductory Physics}.
\newblock Prentice Hall, 2002.

\bibitem{adams2003lecture}
J.~P. Adams, E.~E. Prather, and T.~F. Slater.
\newblock {\em Lecture-Tutorials for Introductory Astronomy}.
\newblock Prentice Hall, Upper Saddle River, NJ, 2005.

\bibitem{beichner2007student}
Robert~J. Beichner, Jeffery~M. Saul, David~S. Abbott, Jeanne~J. Morse, Duane Deardorff, Rhett~J. Allain, Scott~W. Bonham, Melissa~H. Dancy, and John~S. Risley.
\newblock The student-centered activities for large enrollment undergraduate programs {(SCALE-UP)} project.
\newblock {\em Research-based Reform of University Physics}, 1(1):2--39, 2007.

\bibitem{brewe2008modeling}
Eric Brewe.
\newblock Modeling theory applied: Modeling instruction in introductory physics.
\newblock {\em American Journal of physics}, 76(12):1155--1160, 2008.

\bibitem{heller1992teaching}
Patricia Heller, Ronald Keith, and Scott Anderson.
\newblock Teaching problem solving through cooperative grouping.
\newblock {\em American Journal of Physics}, 60(7):627--636, 1992.

\bibitem{heller1992teaching2}
Patricia Heller and Mark Hollabaugh.
\newblock Teaching problem solving through cooperative grouping. part 2: Designing problems and structuring groups.
\newblock {\em American Journal of Physics}, 60(7):637--644, 1992.

\bibitem{hestenes1992force}
David Hestenes, Malcolm Wells, and Gregg Swackhamer.
\newblock Force {C}oncept {i}nventory.
\newblock {\em The Physics Teacher}, 30(3):141--158, 1992.

\bibitem{ramlo2008validity}
Susan Ramlo.
\newblock Validity and reliability of the {Force and Motion Conceptual Evaluation}.
\newblock {\em American Journal of Physics}, 76(9):882--886, 2008.

\bibitem{hestenes1992mechanics}
David Hestenes and Malcolm Wells.
\newblock A {Mechanics Baseline Test}.
\newblock {\em The Physics Teacher}, 30(3):159--166, 1992.

\bibitem{nissen2019missing}
Jayson Nissen, Robin Donatello, and Ben Van~Dusen.
\newblock Missing data and bias in physics education research: A case for using multiple imputation.
\newblock {\em Physical Review Physics Education Research}, 15(2):020106, 2019.

\bibitem{landis1977measurement}
J.~Richard Landis and Gary~G. Koch.
\newblock The measurement of observer agreement for categorical data.
\newblock {\em Biometrics}, pages 159--174, 1977.

\bibitem{dalka2024network}
Robert~P. Dalka and Justyna~P. Zwolak.
\newblock Network analysis of graduate program support structures through experiences of various demographic groups.
\newblock {\em Physical Review Physics Education Research}, 20(2):020106, 2024.

\bibitem{tapping2018visualizing}
Ryan Tapping, G.~P. Lepage, and N.~Holmes.
\newblock Visualizing patterns in {CSEM} responses to assess student conceptual understanding.
\newblock In {\em 2018 Physics Education Research Conference (PERC)}, pages 419--422, 2018.

\bibitem{kortemeyer2014extending}
Gerd Kortemeyer.
\newblock Extending item response theory to online homework.
\newblock {\em Physical Review Special Topics-Physics Education Research}, 10(1):010118, 2014.

\bibitem{odden2020thematic}
Tor Ole~B. Odden, Alessandro Marin, and Marcos~D. Caballero.
\newblock Thematic analysis of 18 years of physics education research conference proceedings using natural language processing.
\newblock {\em Physical Review Physics Education Research}, 16(1):010142, 2020.

\bibitem{tantardini2019comparing}
Mattia Tantardini, Francesca Ieva, Lucia Tajoli, and Carlo Piccardi.
\newblock Comparing methods for comparing networks.
\newblock {\em Scientific Reports}, 9(1):17557, 2019.

\bibitem{steinert2016comparative}
Laura Steinert and H.~Ulrich Hoppe.
\newblock A comparative analysis of network-based similarity measures for scientific paper recommendations.
\newblock In {\em 2016 Third European Network Intelligence Conference (ENIC)}, pages 17--24. IEEE, 2016.

\bibitem{walsh2020connecting}
Cole Walsh, Daniyar Kushaliev, and Natasha~G. Holmes.
\newblock Connecting the dots: Student social networks in introductory physics labs.
\newblock In {\em Physics Education Research Conference}, pages 557--562, 2020.

\bibitem{sundstrom2022examining}
Meagan Sundstrom, David~G. Wu, Cole Walsh, Ashley~B. Heim, and N.~G. Holmes.
\newblock Examining the effects of lab instruction and gender composition on intergroup interaction networks in introductory physics labs.
\newblock {\em Physical Review Physics Education Research}, 18(1):010102, 2022.

\bibitem{foti2011nonparametric}
Nicholas~J. Foti, James~M. Hughes, and Daniel~N. Rockmore.
\newblock Nonparametric sparsification of complex multiscale networks.
\newblock {\em PloS One}, 6(2):e16431, 2011.

\bibitem{rosvall2008maps}
Martin Rosvall and Carl~T. Bergstrom.
\newblock Maps of random walks on complex networks reveal community structure.
\newblock {\em Proceedings of the National Academy of Sciences}, 105(4):1118--1123, 2008.

\bibitem{bruun2014time}
Jesper Bruun and Ian~G. Bearden.
\newblock Time development in the early history of social networks: Link stabilization, group dynamics, and segregation.
\newblock {\em PLoS One}, 9(11):e112775, 2014.

\bibitem{lopez2025mining}
Sonsoles L{\'o}pez-Pernas, Santtu Tikka, and Mohammed Saqr.
\newblock Mining patterns and clusters with transition network analysis: A heterogeneity approach.
\newblock {\em {Advanced Learning Analytics Methods: AI, precision and complexity}}, 2025.

\bibitem{murdoch2002active}
Brock Murdoch and Paul~W. Guy.
\newblock Active learning in small and large classes.
\newblock {\em Accounting Education}, 11(3):271--282, 2002.

\bibitem{gurevitch1999statistical}
Jessica Gurevitch and Larry~V. Hedges.
\newblock Statistical issues in ecological meta-analyses.
\newblock {\em Ecology}, 80(4):1142--1149, 1999.

\bibitem{cohen1992quantitative}
Jacob Cohen.
\newblock Quantitative methods in psychology: A power primer.
\newblock {\em Psychological Bulletin}, 112:1155--1159, 1992.

\bibitem{turner2006calculating}
III Turner, Herbert~M. and Robert~M. Bernard.
\newblock Calculating and synthesizing effect sizes.
\newblock {\em Contemporary issues in communication science and disorders}, 33(Spring):42--55, 2006.

\bibitem{scherr2007enabling}
Rachel~E. Scherr and Andrew Elby.
\newblock Enabling informed adaptation of reformed instructional materials.
\newblock In {\em AIP Conference Proceedings}, volume 883, pages 46--49. American Institute of Physics, 2007.

\bibitem{kinnear2021developing}
George Kinnear, Steph Smith, Ross Anderson, Thomas Gant, Jill~RD MacKay, Pamela Docherty, Susan Rhind, and Ross Galloway.
\newblock Developing the {FILL+} tool to reliably classify classroom practices using lecture recordings.
\newblock {\em Journal for STEM Education Research}, pages 1--23, 2021.

\bibitem{bruun2024semiotic}
Cedric Linder, Jesper Bruun, Arvid Pohl, and Burkhard Priemer.
\newblock Relationship between semiotic representations and student performance in the context of refraction.
\newblock {\em Physical Review Physics Education Research}, 20:010103, 2024.

\bibitem{clauset2004modularity}
Aaron Clauset, M.~E.~J. Newman, and Cristopher Moore.
\newblock Finding community structure in very large networks.
\newblock {\em Phys. Rev. E}, 70:066111, 2004.

\bibitem{mccullough2001gender}
Laura McCullough and David Meltzer.
\newblock Differences in male/female response patterns on alternative-format versions of the {Force Concept Inventory}.
\newblock In {\em Physics Education Research Conference 2001}, PER Conference, Rochester, New York, July 25-26 2001.

\bibitem{han2016experimental}
Jing Han, Kathleen Koenig, Lili Cui, Joseph Fritchman, Dan Li, Wanyi Sun, Zhao Fu, and Lei Bao.
\newblock Experimental validation of the half-length {Force Concept Inventory}.
\newblock {\em Physical Review Physics Education Research}, 12(2):020122, 2016.

\bibitem{singhemcs}
Chandralekha Singh and David Rosengrant.
\newblock Multiple-choice test of energy and momentum concepts.
\newblock {\em American Journal of Physics}, 71(6):607--617, June 2003.

\bibitem{bardar2007development}
Erin~M. Bardar, Edward~E. Prather, Kenneth Brecher, and Timothy~F. Slater.
\newblock Development and validation of the light and spectroscopy concept inventory.
\newblock {\em Astronomy Education Review}, 5(2):103--113, 2007.

\bibitem{slater2014development}
Stephanie~J. Slater.
\newblock The development and validation of the {Test Of Astronomy STandards} {(TOAST)}.
\newblock {\em Journal of Astronomy \& Earth Sciences Education}, 1(1):1--22, 2014.

\end{thebibliography}

\section*{APPENDIX}

\subsection{Comparing Jaccard index and cosine similarity}

\begin{figure}[b]
    \centering
    \includegraphics[width=3.4in,trim={0 0.3cm 0 0}]{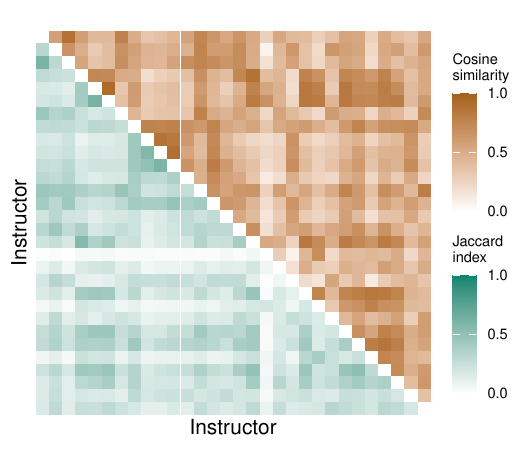}
    \caption{Jaccard index (lower triangle, shades of green) and cosine similarity (upper triangle, shades of brown) values for each pair of classroom observation networks. 
    }
    \label{fig:Heatmap}
\end{figure}

Figure~\ref{fig:Heatmap} shows the Jaccard index and cosine similarity values for each pair of classroom observation networks. As mentioned in the main text, the Jaccard index values (shades of green in the lower triangle) are all relatively low. The cosine similarity values (shades of brown in the upper triangle), however, span many of the intermediate values in the range of 0 to 1.

\subsection{Choosing significance level for sparsifying the similarity network}

\begin{figure}[t]
    \centering
    \includegraphics[width=3.4in,trim={0 0.3cm 0 0}]{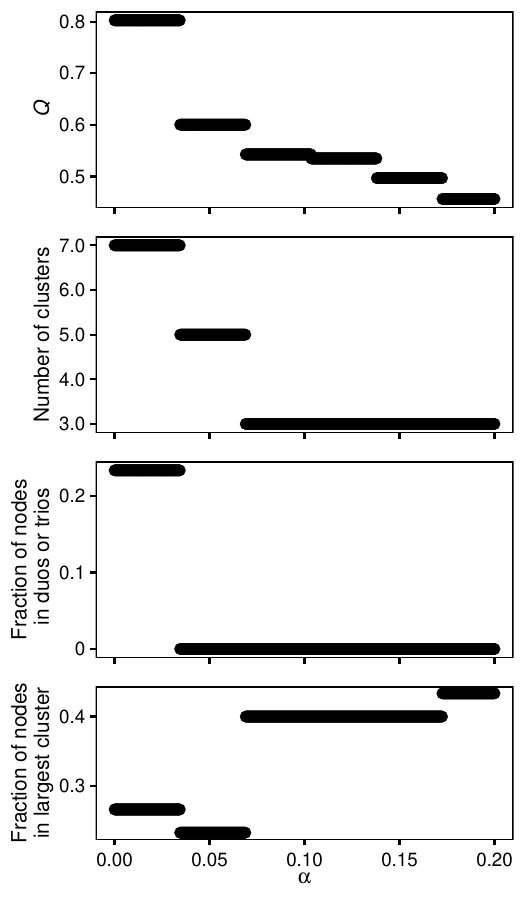}
    \caption{Finding an appropriate significance level, $\alpha$, for network sparsification. All measures were calculated using increments of 0.0001 in $\alpha$. (a) Modularity, $Q$, of the similarity network when it is sparsified with different $\alpha$ values. (b) The number of clusters identified by the Infomap algorithm when the similarity network is sparsified with different  $\alpha$ values. (c) The number of clusters identified by the Infomap algorithm that contain exactly two or three nodes when the similarity network is sparsified with different  $\alpha$ values. (d) The fraction of nodes in the largest cluster identified by the Infomap algorithm when the similarity network is sparsified with different $\alpha$ values.
    }
    \label{fig:Alpha}
\end{figure}

Similar to Ref.~\cite{bruun2024semiotic}, we identified an appropriate significance level, $\alpha$, by
applying the LANS algorithm at $\alpha$ values between 0 and
0.2 in steps of 0.0001 and then running the Infomap community detection algorithm on the resulting sparsified similarity network. We calculated four measures for the Infomap clustering at each $\alpha$ value and chose an $\alpha$ value with suitable properties across all measures (Fig.~\ref{fig:Alpha}).

The first measure, modularity ($Q$), captures the extent to which the sparsified similarity network can be partitioned into meaningful clusters where nodes share more edges with nodes in their cluster than they do with
nodes in other clusters. $Q$ ranges from --0.5 to 1 and $Q > 0.3$ indicates significant modular structure~\cite{clauset2004modularity}. For our similarity network, all $\alpha$ values had $Q > 0.3$ and $\alpha < 0.069$ had $Q > 0.6$ (Fig.~\ref{fig:Alpha}a).

The other three measures helped us to identify $\alpha$ values where the Infomap algorithm extracts multiple different clusters (i.e., not a solution that is one single cluster containing every node) and with multiple nodes in each cluster (i.e., not a solution where some clusters only contain two or three nodes and not a solution where one cluster is much larger than the other clusters). We calculated the number of clusters resulting from the Infomap algorithm at each $\alpha$ value (Fig.~\ref{fig:Alpha}b), the fraction of nodes assigned to clusters containing only two or three nodes (Fig.~\ref{fig:Alpha}c), and the fraction of nodes assigned to the largest cluster (Fig.~\ref{fig:Alpha}d). All $\alpha$ values yielded three or more clusters; however, $\alpha < 0.035$ resulted in some nodes being in clusters of size two or three and $\alpha > 0.069$ resulted in at least 40\% of the nodes being assigned to the largest cluster. Triangulating all of these measures, we decided to use $\alpha = $ 0.05.

\subsection{Distribution of class sizes}

\begin{figure}[t]
    \centering
    \includegraphics[width=3.4in,trim={0 0.3cm 0 0}]{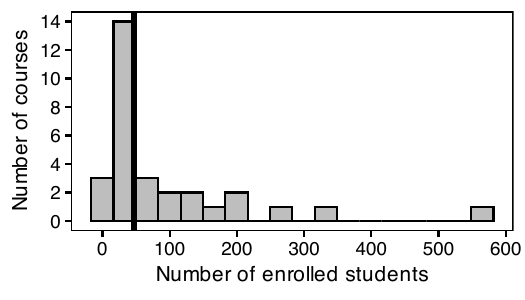}
    \caption{Histogram of class sizes (i.e., number of enrolled students) for the 30 courses included in our study. The thick black line indicates the median of the distribution (48 students).
    }
    \label{fig:ClassSize}
\end{figure}

Figure~\ref{fig:ClassSize} shows the distribution of class sizes for the 30 courses in our study. The median of the distribution is 48 students. We considered courses with more than 45 students to be large and courses with fewer than 45 students to be small because there were no courses with 43 to 47 students and this partition created roughly even numbers of courses in each group: 16 large classes and 14 small classes. This partition is also consistent with that of Freeman and colleagues' meta-analysis, which uses a threshold of 50 students to distinguish between small- and medium-sized classes~\cite{freeman2014active}.

\subsection{Course and institution attributes}

Table~\ref{tab:individualfeatures} provides the course and institution attributes for each individual course included in the study. These attributes were used in our segregration analysis (see Fig.~\ref{fig:seg}) and can be linked to the instruction types we identified in Fig.~\ref{fig:map}.

\begin{table*}[t] 
\centering
\caption{\label{tab:individualfeatures}
Course and institution attributes, using the same course labels as in Fig.~\ref{fig:map}. I-WC indicates whole-class implementation of ISLE, I-LO indicates lab-only implementation of ISLE (i.e., the observation networks represent lecture sections), PI indicates Peer Instruction, T-WC indicates whole-class implementation of Tutorials, T-RO indicates recitation-only implementation of Tutorials (T-RO* indicates recitation-only implementation of Tutorials but with the observation network representing the lecture section of the course), and SU indicates SCALE-UP. SU6 was excluded from our segregation analysis of LPA profiles because each of the three classroom observations were assigned to different profiles, thus we could not assign the course to a profile where the majority of the observations were assigned. FCI indicates Force Concept Inventory~\cite{hestenes1992force}, Gender FCI indicates
Gender Force Concept Inventory~\cite{mccullough2001gender}, Half-FCI indicates
Half-Length Force Concept Inventory~\cite{han2016experimental}, EMCS indicates
Energy and Momentum Conceptual Survey~\cite{singhemcs}, ECA indicates
Energy Concept Assessment, MBT indicates
Mechanics Baseline Test~\cite{hestenes1992mechanics}, FMCE indicates
Force and Motion Concept Evaluation~\cite{ramlo2008validity}, LSCI indicates
Light and Spectroscopy Concept Inventory~\cite{bardar2007development}, and TOAST indicates
Test of Astronomy Standards~\cite{slater2014development}.}
\begin{ruledtabular}
\setlength{\extrarowheight}{1pt}
\begin{tabular}{lccccccc}
Course & LPA profile~\cite{sundstrom2025relativebenefits} & Class size & Public/private & PhD-granting & Discipline & Research designation & Concept inventory\\ 
\hline
I-WC1 & Lecture & 16 (Small) & Private & No & Physics  & Other & FCI\\
I-WC2 & Worksheets & 15 (Small) & Public &  Yes & Physics &  R1 or R2 & FCI\\
I-WC3 & Worksheets & 32 (Small) &  Public & No &  Physics  & Other & FMCE\\
I-WC4 & Worksheets & 33 (Small) &  Public & No  &  Physics  & Other & ...\\
I-LO1 & Lecture & 150 (Large) & Public & Yes & Physics & R1 or R2 & FCI\\
I-LO2 & Lecture & 118 (Large) & Public & Yes & Physics & R1 or R2 & FCI\\
PI1 & Lecture & 28 (Small) & Public& No &  Physics& Other & FCI\\
PI2 & Clickers & 253 (Large) &Public  & Yes & Physics & R1 or R2 & FMCE\\
PI3 & Clickers & 576 (Large) & Private & Yes &  Physics& R1 or R2 & MBT\\
PI4 & Clickers & 201 (Large) & Private & Yes & Physics & R1 or R2 & MBT\\
PI5 & Clickers & 81 (Large) & Private & Yes &  Physics& R1 or R2 & Half-FCI\\
PI6 & Clickers & 94 (Large) & Public & No &  Physics& Other & FCI\\
PI7 & Lecture & 17 (Small) & Private & Yes & Physics & Other & Gender FCI\\
PI8 & Clickers &  320 (Large)& Public & Yes &  Physics& R1 or R2 & FMCE\\
PI9 & Clickers & 20 (Small) &  Private& No &  Physics& Other & FCI\\
T-WC1 & Worksheets & 18 (Small) & Public  & Yes & Astronomy & R1 or R2 & LSCI\\
T-WC2 & Other groupwork & 36 (Small) & Private & Yes & Physics & Other & FMCE\\
T-WC3 & Worksheets & 29 (Small) & Public & No & Physics & Other & FCI\\
T-WC4 & Lecture & 70 (Large) & Private & Yes & Astronomy & R1 or R2 & ...\\
T-RO1 & Worksheets & 32 (Small) &  Private&  Yes& Physics & R1 or R2 & ECA\\
T-RO2 & Worksheets & 49 (Large) &  Public & No & Physics & Other & FCI\\
T-RO3 & Worksheets & 11 (Small) & Public  & No & Physics & Other & FCI\\
T-RO* & Clickers & 185 (Large) & Public & Yes & Astronomy  & R1 or R2 & TOAST\\
SU1 & Worksheets & 48 (Large) & Public &  Yes& Physics & R1 or R2 & ...\\
SU2 & Other groupwork & 54 (Large) & Private & Yes & Physics & R1 or R2 & ...\\
SU3 & Other groupwork & 135 (Large) & Public  & No & Physics & Other & FCI\\
SU4 & Other groupwork &  23 (Small)& Private & No &Physics & Other& FCI\\
SU5 & Lecture & 99 (Large) & Public  & Yes & Physics & R1 or R2& FCI\\
SU6 & ... & 48 (Large) & Private & No  & Physics & Other& FCI\\
SU7 & Worksheets & 42 (Small) & Public & Yes & Physics & R1 or R2 & EMCS\\
\end{tabular} 
\end{ruledtabular}
\end{table*}

\end{document}